\documentclass{tlp}
\usepackage{aopmath}

\usepackage[english]{babel}
\usepackage{harvard}

\newtheorem{example}{Example}

\newlength{\headPlusArrowSpace}
\newcommand{\head}[1]{\ensuremath{\settowidth{\global\headPlusArrowSpace}{$#1 \leftarrow \;$}#1}}
\newcommand{\headWd}{\hskip\headPlusArrowSpace}

\newcounter{contador}

\newcommand {\nil} {[\,]}
\newcommand {\var} {\mbox{var}}
\newcommand {\Var} {{\rm var}}

\newcommand {\demo} {\mbox{\it demo}}
\newcommand {\True} {\mbox{\it true}}
\newcommand {\Axiom} {\mbox{\it axiom}}
\newcommand {\singleCallSet} {\mbox{\it single\_call\_set}}
\newcommand {\allSet} {\mbox{\it all\_set}}

\newcommand {\allMatches} {\mbox{\it all\_matches}}

\newcommand {\Copy} {\mbox{\it copy}}
\newcommand {\termInstances} {\mbox{\it term\_instances}}
\newcommand {\Definition} {\mbox{\it definition}}
\newcommand {\unify} {\mbox{\it unify}}
\newcommand {\Set} {\mbox{\it set}}
\newcommand {\ap} {\mbox{\it ap}}
\newcommand {\append} {\mbox{\it append}}
\newcommand {\unit} {\mbox{\it unit}}
\newcommand {\isunit} {\mbox{\it is\_unit}}
\newcommand {\nonunit} {\mbox{\it nonunit}}
\newcommand {\Defn} {\mbox{\it defn}}
\newcommand {\Union} {\mbox{\it union}}
\newcommand {\Not} {\mbox{\it not}}
\newcommand {\naiveh} {\mbox{\it naive\_h}}
\newcommand {\Rename} {\mbox{\it rename}}
\newcommand {\Sub} {\mbox{\it Sub}}
\newcommand {\match} {\mbox{\it match}}
\newcommand {\apply} {\mbox{\it apply}}
\newcommand {\goon} {\mbox{\it halt\_cont}}

\newcommand {\Read} {\mbox{\it read}}
\newcommand {\Write} {\mbox{\it write}}
\newcommand {\More} {\mbox{\it '\ more?\ '}}

\newcommand {\Halt} {\mbox{\it HaltCont}}
\newcommand {\halt} {\mbox{\it halt}}
\newcommand {\cont} {\mbox{\it cont}}
\newcommand {\noans} {\mbox{\it no\_ans}}
\newcommand {\ans} {\mbox{\it ans}}
\newcommand {\stream} {\mbox{\scriptsize\it stream}}
\newcommand {\And} {\mbox{\sc and}}
\newcommand {\Or} {\mbox{\sc or}}
\newcommand {\allq} {\mbox{\it all\_q}}
\newcommand {\x} {{\bf x}}

\newcommand {\ys} {{\bf ys}}
\newcommand {\Ws} {\mbox{\em Ws}}
\newcommand {\Xs} {\mbox{\em Xs}}
\newcommand {\Ys} {\mbox{\em Ys}}
\newcommand {\Zs} {\mbox{\em Zs}}
\newcommand {\YsZs} {\mbox{\em YsZs}}
\newcommand {\St} {\mbox{\em St}}
\newcommand {\hs} {{\hat{s}}}
\newcommand {\hats} {{\hat{s}}}
\newcommand {\rhs} {\mbox{rhs}}
\newcommand {\ha} {{\hat{a}}}
\newcommand {\StXs} {\mbox{\em StXs}}
\newcommand {\StYs} {\mbox{\em StYs}}
\newcommand {\StZs} {\mbox{\em StZs}}
\newcommand {\StYsZs} {\mbox{\em StYsZs}}

\newcommand {\Qs} {\mbox{\em Qs}}
\newcommand {\Pjs} {\mbox{\em Pjs}}
\newcommand {\Pj} {\mbox{\em Pj}}
\newcommand {\QsQs} {\mbox{\em QsQs}}
\newcommand {\Bs} {\mbox{\em Bs}}
\newcommand {\ABs} {{A\! B\hspace{-1.4pt}s}}
\newcommand {\Atheta} {{A\theta}}
\newcommand {\Athetas} {{A\theta\hspace{-1pt}s}}
\newcommand {\ABthetas} {{A\! B\hspace{-0.5pt}\theta\hspace{-1pt}s}}
\newcommand {\Bthetas} {{B\hspace{-0.5pt}\theta\hspace{-1pt}s}}
\newcommand {\Btheta} {{B\hspace{-0.5pt}\theta}}
\newcommand {\Cthetas} {{C\hspace{-1pt} \theta\hspace{-1pt}s}}
\newcommand {\Ctheta} {{C\hspace{-1pt} \theta}}
\newcommand {\Xtheta} {{X\hspace{-1pt} \theta}}

\newcommand {\identityRel} {\mbox{\em I}}

\def\igualdef{\; \buildrel {..} \over = \;}

\def\comp{\mathbin{\mbox{$;$}}}

\def\Insert{\mathbin{\mbox{$\diamond$}}}

\begin{document}
\bibliographystyle{dcu}

\title[Chain Programs for Writing Deterministic Metainterpreters]
{Chain Programs for Writing \\ Deterministic Metainterpreters}
\author[David A.\ Rosenblueth]
{DAVID A.\ ROSENBLUETH \\
Instituto de Investigaciones en Matem\'aticas Aplicadas y en Sistemas \\
Universidad Nacional Aut\'onoma de M\'exico \\
Apdo.\ 20-726, 01000 M\'exico D.F.}

\maketitle

\begin{abstract}
Many metainterpreters found in the logic programming
literature are {\em nondeterministic} in the sense that 
the 
selection of program clauses
is not determined.
Examples are the familiar ``demo'' and ``vanilla''
metainterpreters.
For some applications this nondeterminism is 
convenient. 
In some cases, however, a {\em deterministic} metainterpreter, having an
explicit selection of clauses, is needed.
Such cases include 
(1)~conversion of \Or\ parallelism into \And\ parallelism for
``committed-choice'' processors,
(2)~logic-based, imperative-language implementation of search strategies,
and
(3)~simulation of bounded-resource reasoning.

Deterministic metainterpreters are difficult to write because
the programmer must be concerned about the set of unifiers 
of the children of a node in the derivation tree.
We argue that it is 
both possible and advantageous
to write these metainterpreters by reasoning 
in terms of
object programs converted into a syntactically restricted form that
we call ``chain'' form,
where we can forget about unification, except for unit clauses.
We give two transformations converting logic programs into chain form,
one for ``moded'' programs (implicit in two existing
exhaustive-traversal methods for committed-choice execution), and one
for arbitrary definite programs.
As illustrations of our approach we show examples of the
three applications mentioned above.
\end{abstract}

\section{Introduction}
\label{ideas}

Perhaps the most common use of metalogic is the definition and
implementation of metainterpreters~\cite{ssh86a,abr88a,kwl90,atr95}.
Many applications of metainterpreters
are based on concise definitions, like that of
the ``vanilla'' metainterpreter, which can be easily elaborated as
required.
Other applications, however, have been neglected, possibly because of
employing convoluted definitions. 
Examples are deterministic metainterpreters exhaustively traversing
search spaces.
Our purpose will be to present a technique simplifying the
design of deterministic metainterpreters.
This technique converts the object program into a form severely
restricted in its syntax, thereby facilitating reasoning about its
search space.

Early works exploiting 
metainterpreters to great
advantage are for example Bowen and Kowalski's amalgamation of
language and metalanguage~\cite{bkw82}, Sergot's ``query-the-user''
facility~\cite{srg82}, and Shapiro's ``algorithmic debugger''~\cite{shp82}.
These metainterpreters, just as the familiar vanilla and demo
metainterpreters, have 
nondeterministic definitions.
Consider for example the 
following
{\it demo} predicate~\cite{kwl90,kwl95}:
\begin{eqnarray*}
&& \demo(T,P) \leftarrow \Axiom(T,P\leftarrow Q),\ \demo(T,Q) \\
&& \demo(T,P \wedge Q ) \leftarrow \demo(T,P),\ \demo(T,Q) \\
&& \demo(T,\True) \leftarrow
\end{eqnarray*}
This definition is nondeterministic because it is not
determined, in the first clause of the definition, 
what axiom in the theory $T$, having conclusion $P$,
might be needed to demonstrate $P$~\cite[p.\ 229]{kwl95}.

For some applications this nondeterminism is convenient.
For others, however, a {\em deterministic} metainterpreter, having an
explicit selection of axioms, is desired.
A problem amenable to deterministic metainterpretation is
that of exhaustively traversing, using a committed-choice processor,
the search space generated by a logic
program and a goal~\cite{ud87,tmk87}.
Another problem 
is that of describing search strategies with logic
programs equivalent to flowcharts,
like the programs of~\cite{cvn81}.
Yet another application is
Kowalski's approach for reconciling reactive and rational agents with
bounded-resource metainterpreters~\cite{kwl95}.

It is of course possible to write deterministic metainterpreters for
logic programs.
Clark and Gregory were perhaps among the first to publish~\cite{cgr85}
one such metainterpreter.
We show a slightly modified version of their metainterpreter in
Fig.~\ref{clgr}.
The intended meanings of some of the predicates in this metainterpreter
are as follows.
Assume that the set of answer substitutions to
the goal having $A$ as its only subgoal is
$\{\theta_1,\ldots,\theta_n\}$.
Then the predicate $\singleCallSet( A, \Athetas )$ is intended to hold when
$\Athetas$ is a list of the form
$[A\theta_1,\ldots,A\theta_n]$.

The predicate $\Set( A, \Bs, \Athetas )$ is a generalisation of
$\singleCallSet$.
In this case the goal $\Bs$, which may have more than one subgoal, is of
the form $B_1 \wedge \ldots \wedge B_m \wedge \True$, and $\Athetas$ is 
of the form
$[A\theta_1,\ldots,A\theta_n]$, 
where $\{\theta_1,\ldots,\theta_n\}$
is the set of answer substitutions for $\Bs$.
($\Bs$ terminates by $\True$ to simplify the code.)

The predicate
$\allSet( \ABs, \Athetas )$, in turn, can be viewed as a generalisation of
$\Set$, where $\ABs$ is a list of clauses 
$[A_1\leftarrow\Bs_1,\ldots,A_m\leftarrow\Bs_m]$ and $\Athetas$ is 
of the form $[A_1\theta_{1,1},\ldots,A_1\theta_{1,n_1},\ldots,
A_m\theta_{m,1},\ldots,A_m\theta_{m,n_m}]$, 
where $\{\theta_{i,1},\ldots,\theta_{i,n_i}\}$
is the set of answer substitutions for $\Bs_i$.
We refer the reader to~\cite{cgr85} for a thorough discussion of
this metainterpreter.

As exhibited in Fig.~\ref{clgr}, the programmer must be concerned
about {\em the set of unifiers of the children of a node in the
derivation tree} 
(cf.\ the $\termInstances$ predicate).
The reason for this concern, as we will see,
is that variables in a logic program 
can appear anywhere in a clause.
This is an additional difficulty, absent in usual nondeterministic
metainterpreters.

\begin{figure}[hbt]
\small
\figrule
Example of object-program representation:
\begin{eqnarray*}
&& \Definition( \ap(\_,\_,\_), [\ap(\nil,L,L) \leftarrow \True, \\
&& \hspace{2.96cm} 
\ap([A|L'],M,[A|N]) \leftarrow ( \ap(L',M,N) \wedge
\True ) \\
&& \hspace{2.9cm}
] \\
&& \hspace{1.4cm} ) 
\leftarrow 
\end{eqnarray*}
Example of goal:
$\leftarrow \singleCallSet( \ap(X,Y,[a,b]), \Athetas )$

\vspace{15pt}

Metainterpreter:
\begin{eqnarray*}
&& \head{\singleCallSet( A, \Athetas )} \leftarrow
        \Definition( A, \ABs ), \\
&& \headWd
        \allMatches( A, \ABs, \ABthetas' ), \\
&& \headWd \allSet( \ABthetas', \Athetas ) \\
&& \allSet( \nil, \nil ) \leftarrow \\
&& \head{\allSet( [A\leftarrow\Bs|\ABs], \Athetas )} \leftarrow
      \Set( A, \Bs, \Athetas_1 ), \\
&& \headWd \allSet( \ABs, \Athetas_2 ), \\
&& \headWd \append( \Athetas_1, \Athetas_2, \Athetas ) \\
&& \Set( A, \True, [A] ) \leftarrow \\
&& \head{\Set( A, B\wedge\Bs, \Athetas )} \leftarrow
	\singleCallSet( B, \Bthetas ), \\
&& \headWd
    \termInstances( B, \Bthetas, A\leftarrow\Bs, \ABthetas ), \\
&& \headWd \allSet( \ABthetas, \Athetas ) \\
&& \allMatches( A, \nil, \nil ) \leftarrow \\
&& \head{\allMatches( A, [\Atheta\leftarrow\Bthetas|\ABs], 
                   [\Atheta\leftarrow\Bthetas|\ABthetas'] )} \leftarrow 
       \Copy( A, \Atheta' ), \\
&& \headWd \unify( \Atheta, \Atheta' ),\ !, \\
&& \headWd \allMatches( A, \ABs, 
                              \ABthetas' ) \\
&& \allMatches( A, [A'\leftarrow\Bs|\ABs], 
                                       \ABs' ) \leftarrow 
\allMatches( A, \ABs, \ABs' ) \\
&& \termInstances( B, \nil, C, \nil ) \leftarrow \\
&& \head{\termInstances( B, [\Btheta|\Bthetas], C, [\Ctheta|\Cthetas] )}
       \leftarrow \Copy( f(B,C), f(\Btheta',\Ctheta) ), \\
&& \headWd \unify( \Btheta, \Btheta' ), \\
&& \headWd \termInstances( B, \Bthetas, C, \Cthetas )
\end{eqnarray*}
\caption{\it A deterministic, exhaustive-traversal metainterpreter
for arbitrary definite programs.
For readability, we use a string $X\hspace{-1.2pt}\theta$ 
as the name of a variable taking as value an
instance of the value of $X$.
Similarly, $X\hspace{-1.2pt}\theta\hspace{-1pt}s$
is the name of a variable taking as value a list of instances of
the value of $X$.}
\label{clgr}
\figrule
\end{figure}

The study of deterministic metainterpreters can be viewed
as an attempt to narrow the ``gap'' between the ``don't-know'' form of
logic programming, needed for user-level applications, and the 
``don't-care'' form, useful for controlling the execution of logic
programs.
Kowalski's observation~\cite{kwl93a}
that the FGCS project had experienced such a gap
suggests looking at the work done in connection
with this project.

First Ueda~\cite{ud87}, and then Tamaki~\cite{tmk87}, 
published methods converting a
nondeterministic logic program into a deterministic version.
The motivation for developing such methods was that of allowing
the execution of \Or-parallel programs by committed-choice processors
(which are \And\ parallel).
Understanding how these methods work might lead to key ideas for obtaining
other deterministic-evaluation methods.
However, the considerable intricacy of these methods is an
obstacle for giving a clear and concise explanation of their central
mechanisms. 

Trying to elucidate the principles on which these methods
are based, we found
an important common characteristic.
Both 
tend to hide, as it were, certain occurrences
of variables, in a list behaving like a stack.
In particular, such occurrences are those of variables that 
(1)~receive a substitution {\em before} a subgoal $A$ is selected
and (2)~occur in a subgoal selected {\em after} $A$ has succeeded.
Variables having such occurrences
are called ``pass-on'' variables in~\cite{tmk87}.
Hence, {\em if we consider programs lacking pass-on variables, a fortiori,
these transformations get simplified}.

A kind of program lacking
pass-on variables is that of ``chain'' programs, having clauses of the
form:
\begin{eqnarray*} 
&& p(X_0,X_n) \leftarrow q_1(X_0,X_1),\ q_2(X_1,X_2),\ \ldots,\ 
                         q_n(X_{n-1},X_n) 
\end{eqnarray*}
(as well as other clauses; we formally define chain programs in
Sect.~\ref{meta:int:chain}). 
Apparently, concentrating on chain programs should simplify
the task of devising deterministic-traversal methods in general, and 
defining deterministic metainterpreters in particular.
We confirm this possibility by deriving
in Sect.~\ref{meta:int:chain} a deterministic
metainterpreter by reasoning in terms of relational union and
composition.

This metainterpreter is more useful once we 
provide ways of transforming logic programs that do not have chain
form into such a form.
The methods of~\cite{ud87} and~\cite{tmk87} can handle
``moded'' programs, that do not necessarily have chain form.
In these programs,
each argument place of each predicate 
is used either as input (instantiated) or as output (uninstantiated).
By comparing the original methods with their 
versions simplified to chain programs, we have uncovered
a transformation converting moded programs into chain form. 
In Sect.~\ref{sect:moded} we give such a transformation, which is in
fact implicit in~\cite{ud87} and~\cite{tmk87}.
As in the methods of Ueda and Tamaki, we {\em hide pass-on variables in
a list behaving like a stack}.
(We have previously used this transformation for adapting parsers for
context-free languages obtaining inference
systems for moded logic programs~\cite{rsn96,rpr96}.)

Next, in Sect.~\ref{sect:definite} we give another transformation,
converting arbitrary definite programs into chain form,
inspired by the previous one.

Once we have this more general transformation, we can easily extend,
in Sect.~\ref{reconstr}, the existing methods for deterministic,
exhaustive traversal~\cite{ud87,tmk87} to handle arbitrary
definite programs.

Section~\ref{other:app} shows how to write deterministic
metainterpreters for other applications.
First we exhibit a metainterpreter having the same behaviour as that
of Prolog systems.
Next we give a bounded-resource metainterpreter.

We will assume some familiarity with logic program transformation 
through the unfold/fold rules~\cite{ppr94}.

Code appearing throughout the sequel, as well as further examples
of programs converted into chain form, may be found
at:
\begin{verse}
\verb|http://leibniz.iimas.unam.mx/~drosenbl/detmeta|.
\end{verse}

\section{A Deterministic, Exhaustive-Traversal Metainterpreter for
Chain Programs} 
\label{meta:int:chain}

In this section, we will write a 
deterministic, exhaustive-traversal metainterpreter for chain programs
by reasoning in terms of relational union and composition.
A chain program can be viewed as defining a system of equations of
relational expressions.
Such systems of relational equations have been studied 
for example in~\cite{eng74,bmr75,blk77a}.
Our translation of chain programs into relational equations
enables us to regard logical inference from such programs
as the evaluation of relational expressions built from union and
composition.
Our chain programs are similar to, but different from, certain programs
occurring in the deductive-database literature under the same name;
a difference is that we allow function symbols other than constants.

For clarity, we will discuss now this metainterpreter assuming that
only ground terms are constructed.
However, in Sect.~\ref{sect:definite} we will observe that with
the addition of variable renaming and
full unification (as opposed to matching, i.e.\
one-way unification),
this metainterpreter is also valid for constructing terms with
variables.

\subsection{Chain programs as systems of relational equations}

We define a {\em chain program} as a logic program consisting only of
clauses of the form:
\begin{eqnarray} 
&& p(X_0,X_n) \leftarrow q_1(X_0,X_1),\ q_2(X_1,X_2),\ \ldots,\ 
                         q_n(X_{n-1},X_n) \hspace{0.3cm} n > 0
                         \label{chain:cl} \\
\mbox{or } && p(t,t') \leftarrow \label{prim:cl}
\end{eqnarray}
where the $X_i$'s are distinct variables, and $t$ and $t'$ are any term.
The first argument place of a predicate will be called its {\em
input}, and the second argument place its {\em output}.
It will be useful to single out a kind of chain program where all
answers for such a program and a goal with a leftmost ground input are
ground, assuming a leftmost computation rule.
If $\var(t') \subseteq \var(t)$ in every clause of the
form~(\ref{prim:cl}), then the program is called a {\em G-chain
program}.
Here, and throughout the sequel, we use $\var(t)$ to denote the set of
variables in expression $t$.

Clearly, the clause~(\ref{chain:cl}) denotes the inclusion:
\begin{eqnarray}
&& P \supseteq (Q_1 \comp Q_2 \comp \ldots \comp Q_n) \label{pqs}
\end{eqnarray}
where $P$, $Q_i$ name the relations denoted by $p$, $q_i$,
respectively, and ``;'' denotes relational 
composition.\footnote{The composition of relations $P$ and $Q$ is
defined as: $P \comp Q \igualdef \{ (x,z) : \exists y [ 
(x,y) \in P \; \& \; (y,z) \in Q ] \}$.}

Let us define 
\begin{eqnarray*}
&& P_j \igualdef Q_1 \comp Q_2 \comp \ldots \comp Q_n
\end{eqnarray*}
if the $j$-th clause defining the predicate with symbol
$p$ has the form~(\ref{chain:cl}).
If, on the other hand,
the $j$-th clause defining the predicate with symbol $p$ 
has the form~(\ref{prim:cl}), then
\begin{eqnarray*}
&& P_j \igualdef 
\{ (\x,\x') : p(\x,\x') \leftarrow
              \mbox{ is a ground instance of $p(t,t') \leftarrow$} \}
\end{eqnarray*}

Hence, a chain program denotes a system of relational expressions having
an inclusion of the form
\begin{eqnarray*}
&& P \supseteq (P_1 \cup P_2 \cup \ldots \cup P_m)
\end{eqnarray*}
for each largest set of clauses defining a predicate with symbol
$p$. 
Just as sometimes the meaning of a logic program is defined as its least
Herbrand model, here
we are interested in the least solution of this system.
It can be shown that such a solution is
equal to the unique solution of the system obtained by replacing the
inclusions by equalities.

For notational convenience,
we will extend relational composition to the
case where the first argument is a set.
Let $S \subseteq D$ and $R \subseteq D \times D$.
\begin{eqnarray*}
&& S \comp R \igualdef \{ z : \exists y [ 
y \in S \;\&\; (y,z) \in R ] \}
\end{eqnarray*}
i.e.\ $S \comp R$ denotes the image of $S$ under $R$.
Also, we will sometimes omit the curly braces in singletons.
By $\identityRel$ we will denote the identity relation and by
$\emptyset$ the empty relation as well as the empty set.

Given an object chain program,
we will use $X$ as a metavariable taking as value a ground term,
$\Xs$ a set of ground terms,
$Q$ a relation denoted by an (object-level) predicate, 
$\Qs$ compositions $Q_1 \comp \ldots \comp Q_n$ of relations
denoted by predicates,
$\Pj$ a relation denoted by a single clause, and
$\Pjs$ unions $P_1 \cup \ldots \cup P_m$ of relations denoted 
each by a single clause.

Computing all answers for a chain program and a goal $\leftarrow
q(\x,Z)$, where $\x$ is a ground term,
translates to evaluating the expression
$\x \comp Q$, where $Q$ names the relation denoted by $q$.
During this evaluation, we will have to evaluate relational
expressions of the form:
\begin{eqnarray}
&& \Xs \comp \Qs \label{composition} 
\end{eqnarray}
which represent \And\ branches of the SLD tree.

\subsection{A metainterpreter as a relational-expression evaluator}
Let us now establish an object-program representation.
For simplicity, we use the ambivalent syntax of~\cite{jng94}.
If the $j$-th clause defining a predicate with symbol $p$ is of the
form~(\ref{chain:cl}) 
we will have the following clause at the metalevel:
\begin{eqnarray*}
&& \nonunit(p_j,(q_1 \comp q_2 \comp \ldots \comp q_n \comp \identityRel))
\leftarrow
\end{eqnarray*}
where ``$;$'' can be interpreted
as a right-associative infix list constructor and $\identityRel$ as a
constant. 
If the $j$-th clause defining a predicate with symbol $p$ is of the
form~(\ref{prim:cl})
we will have:
\begin{eqnarray*}
&& \isunit(p_j) \leftarrow \hspace{0.5cm} \mbox{ and} \\
&& \unit(p_j,t,t') \leftarrow 
\end{eqnarray*}
(The $\isunit$ predicate is clearly unnecessary, as could be defined using
the $\unit$ predicate, but we will use it for readability.)

Also, for each largest set of clauses defining a predicate with symbol
$p$ we will have:
\begin{eqnarray*}
&& \Defn(p, (p_1 \cup p_2 \cup \ldots \cup p_m \cup
\emptyset)) \leftarrow 
\end{eqnarray*}
where $\cup$ can be interpreted
as a right-associative infix list constructor denoting set union
and $\emptyset$ as a constant. 

In general, we could have two main evaluation strategies
for~(\ref{composition}):
``termwise''
(i.e.\ decomposing $\Xs$)
and ``relationwise'' 
(i.e.\ decomposing $\Qs$).
Here we will concentrate on a termwise definition (but see~\cite{rsn98a}
for an application of a metainterpreter using a relationwise definition).

Consider~(\ref{composition}).
The predicate $a(\Xs,\Qs,\Zs)$ is intended to hold if $\Zs$ is $\Xs
\comp \Qs$.
Then the next two clauses, which constitute a termwise definition of
this predicate, 
follow from the distributivity of composition over union:
\begin{eqnarray*}
&& a( \emptyset, \Qs, \emptyset ) \leftarrow \nonumber \\
&& \head{a( \{X\} \cup \Xs, \Qs, \YsZs )} \leftarrow
	a'( X, \Qs, \Ys ),\ 
	a( \Xs \setminus \{X\}, \Qs, \Zs ), \\ \nonumber
&& \headWd \Union( \Ys, \Zs, \YsZs )
\end{eqnarray*}
where $a'(X,\Qs,\Ys)$ is meant to hold when $\Ys$ is $X \comp \Qs$
and $\Union(\Ys,\Zs,\YsZs)$
is meant to hold when $\YsZs$ is $\Ys \cup \Zs$.

We can decompose $\Qs$ in the definition of $a'$, which follows from
the definition of composition:
\begin{eqnarray*}
&& a'( X, I, \{X\} ) \leftarrow \nonumber \\
&& a'( X, (Q \comp \Qs), \Zs ) \leftarrow
	b'( X, Q, \Ys ),\ 
	a( \Ys, \Qs, \Zs )
\end{eqnarray*}
where $b'(X,Q,\Ys)$ represents the composition $\Ys$ of a single
term $X$ with a single relation $Q$.

The following clause translates 
such a composition
into the composition of $X$ with a union $\Pjs$ of relations:
\begin{eqnarray*}
&& b'( X, Q, \Ys ) \leftarrow 
	\Defn( Q, \Pjs ),\
	c'( X, \Pjs, \Ys ) \nonumber
\end{eqnarray*}
where $c'(X,\Pjs,\Ys)$ is intended to hold if $\Ys$ is $X \comp \Pjs$.

Now we inductively define $c'$ by decomposing $\Pjs$.
This definition of $c'$, as that of $a$, follows from 
the distributivity of composition over union:
\begin{eqnarray*}
&& c'( X, \emptyset, \emptyset ) \leftarrow \nonumber \\
&& \head{c'( X, \{\Pj\} \cup \Pjs, \YsZs )} \leftarrow
	d'( X, \Pj, \Ys ),\ 
	c'( X, \Pjs \setminus \{\Pj\}, \Zs ), \\ \nonumber
&& \headWd \Union( \Ys, \Zs, \YsZs ) 
\end{eqnarray*}
where the predicate $d'(X,\Pj,\Ys)$ is assumed to hold when $\Ys$
is $X \comp \Pj$, and the predicate $\Union(\Ys,\Zs,\YsZs)$
is assumed to hold when $\YsZs$ is $\Ys \cup \Zs$.

Next we write a definition of $d'(X,\Pj,\Ys)$.
This definition
uses an auxiliary predicate $e'(X,\Pj,\Ys)$ in case $\Pj$ represents
a unit clause.
If, on the other hand, $\Pj$ represents a nonunit clause, then the $d'$
predicate uses the (object-level) definition of $\Pj$ to translate $X
\comp \Pj$ into 
$X \comp \Qs$, so that the previously defined predicate 
$a'( X, \Qs, \Zs )$
can be used.
\begin{eqnarray*}
&& d'( X, \Pj, \Ys ) \leftarrow 
	\isunit( \Pj ),\ 
        e'( X, \Pj, \Ys ) \nonumber \\
&& d'( X, \Pj, \Zs) \leftarrow
	\nonunit( \Pj, \Qs ),\ 
	a'( X, \Qs, \Zs )
\end{eqnarray*}

It only remains to define
$e'(X,\Pj,\Ys)$, intended to hold if $\Ys$ is $X \comp \Pj$ and
$\Pj$ represents a unit clause.
\begin{eqnarray}
&& e'( X, \Pj, \{Y\} ) \leftarrow 
	\unit( \Pj, X, Y ) \label{match} \\
&& e'( X, \Pj, \emptyset ) \leftarrow
	\Not( \unit( \Pj, X, \_ ) ) \label{nomatch}
\end{eqnarray}
The clause~(\ref{match}) covers the case where $X$ matches the input of
the unit clause named $\Pj$, whereas the clause~(\ref{nomatch}) covers the 
case where $X$ does not
match the input of the unit clause named $\Pj$.

Finally, we give
the complete metainterpreter in a more
standard Prolog notation.
In addition, we have approximated set union with list concatenation.
(A more efficiently executable metainterpreter would use difference
lists, but for clarity we prefer ordinary lists.)
We will call this the {\em abcde} metainterpreter.
\begin{eqnarray}
&& a( \nil, \Qs, \nil ) \leftarrow \nonumber \\
&& \head{a( [X|\Xs], \Qs, \YsZs )} \leftarrow
	a'( X, \Qs, \Ys ),\ 
	a( \Xs, \Qs, \Zs ), \nonumber \\
&& \headWd \append( \Ys, \Zs, \YsZs ) \label{abcd:a} \\
&& \ \nonumber \\
&& a'( X, \nil, [X] ) \leftarrow \nonumber \\
&& a'( X, [Q|\Qs], \Zs ) \leftarrow
	b'( X, Q, \Ys ),\ 
	a( \Ys, \Qs, \Zs ) \label{abcd:a1} \\
&& \ \nonumber \\
&& b'( X, Q, \Ys ) \leftarrow 
	\Defn( Q, \Pjs ),\
	c'( X, \Pjs, \Ys ) \nonumber \\
&& \ \nonumber \\
&& c'( X, \nil, \nil ) \leftarrow \nonumber \\
&& \head{c'( X, [\Pj|\Pjs], \YsZs )} \leftarrow
	d'( X, \Pj, \Ys ),\ 
	c'( X, \Pjs, \Zs ),\nonumber \\
&& \headWd \append( \Ys, \Zs, \YsZs ) \label{abcd:c} \\
&& \ \nonumber \\
&& d'( X, \Pj, \Ys ) \leftarrow 
	\isunit( \Pj ),\ 
        e'( X, \Pj, \Ys ) \nonumber \\
&& d'( X, \Pj, \Zs) \leftarrow
	\nonunit( \Pj, \Qs ),\ 
	a'( X, \Qs, \Zs ) \label{abcd:d} \\
&& \ \nonumber \\
&& e'( X, \Pj, [Y] ) \leftarrow 
	\unit( \Pj, X, Y ) \nonumber \\
&& e'( X, \Pj, \nil ) \leftarrow
	\Not( \unit( \Pj, X, \_ ) ) \nonumber
\end{eqnarray}

\section{Conversion of Moded Programs into Chain Form}
\label{sect:moded}

Having written a deterministic metainterpreter for chain programs, our
aim now is to develop
a transformation converting 
``moded'' programs~\cite{apt97} into chain form.
We will derive such a transformation with unfold/fold rules~\cite{ppr94}.
This transformation is in fact implicit in two existing methods for
deterministic, exhaustive traversal: the continuation-based~\cite{ud87}
and the stream-based~\cite{tmk87} methods.
We have previously used such a transformation for adapting parsers for
context-free grammars obtaining inference
systems for moded logic programs~\cite{rsn96,rpr96}.

A clause
\begin{eqnarray}
&& p(t_0,t_n') \leftarrow q_1(t_0',t_1),\ q_2(t_1',t_2),\ 
                    \ldots,\ q_n(t_{n-1}',t_n)
\hspace{1cm} n \geq 0 \label{moded}
\end{eqnarray}
is called {\em moded} if:
\begin{enumerate}
\item $\var(t_i') \subseteq \var(t_0) \cup \cdots \cup \var(t_i)$,
for $i=0,\ldots,n$; and
\item $\var(t_i) \cap \var(t_j) = \emptyset$,
for $i,j=0,\ldots,n$ and $i \not = j$.
\end{enumerate}
A program is called {\em moded} if it consists
only of moded clauses.

Condition~1 causes every input to be ground if the input of the
initial goal is also ground and we use a leftmost computation rule.
When a subgoal succeeds,
condition~2 causes 
the constructed term to
have an effect only on the input of other subgoals, thus avoiding
speculative bindings.
In~\cite{rsn96,rpr96} we had a third 
condition (that
each variable occurring in $t_i'$ occurs only once
in $t_i'$, for $i=0,\ldots,n$, if $n>0$)
meant only for simplifying the transformation.
However, as we observe below,
it is possible to eliminate such a condition without 
excessively elaborating the transformation.

We will use the {\em standard equality theory}.
Given a program $P$,
this theory consists of the following axioms:
\begin{eqnarray*}
&& X=X \leftarrow \\
&& X=Y \leftarrow Y=X \\
&& X=Z \leftarrow X=Y, \ Y=Z \\
&& \{ f(X_1,\ldots,X_{n_f})=f(Y_1,\ldots,Y_{n_f}) \leftarrow 
                          X_1=Y_1,\ \ldots,\ X_{n_f}=Y_{n_f} : \\
&& \hspace{5cm} f \mbox{ is a function symbol occurring in } P \} \\
&& \{ p(X_1,\ldots,X_{n_p}) \leftarrow X_1=Y_1,\ \ldots,\ X_{n_p}=Y_{n_p},\ 
                 p(Y_1,\ldots,Y_{n_p}) : \\
&& \hspace{5cm} p \mbox{ is a predicate symbol occurring in } P \} 
\end{eqnarray*}
which are called, reflexivity, symmetry, transitivity,
function substitutivity, and predicate substitutivity, respectively.

One way of obtaining a chain clause from a moded clause
would be to resolve first the moded clause
with predicate substitutivity so as to
replace each argument of a subgoal by a variable
and then to fold the resulting clause using
some new predicates so as to remove the introduced 
equations.
We will see, however, that such a folding operation may not always be
sound~\cite{gsh92,tst84}.

Consider the following $\append$ program used for splitting lists.
\begin{eqnarray}
&& s(\langle L \rangle, \langle \nil, L \rangle) \leftarrow
\label{s1} \\
&& \underbrace{\rule[-2.5pt]{0pt}{1cm}s}_{\mbox{$p$}}(
        \underbrace{\langle [A|N] \rangle}_{\mbox{$t_0$}}, 
        \underbrace{\langle [A|L],M \rangle}_{\mbox{$t_1'$}}) 
\leftarrow 
   \underbrace{\rule[-2.5pt]{0pt}{1cm}s}_{\mbox{$q_1$}}( 
        \underbrace{\langle N \rangle}_{\mbox{$t_0'$}}, 
        \underbrace{\langle L,M \rangle}_{\mbox{$t_1$}} ) \label{s2}
\end{eqnarray}
We employ angled brackets $\langle\;\rangle$ instead of
ordinary brackets $[\;]$ for grouping input and output arguments.
We do this for clarity.

By predicate substitutivity and symmetry, it is possible to derive
from~(\ref{s2}): 
\begin{eqnarray}
&& \head{s(X,Y)} \leftarrow \fbox{$X=\langle [A|N] \rangle$},\ 
                     Y=\langle [A|L],M \rangle,\ 
                     \fbox{$X'=\langle N \rangle$},\ 
                     Y'=\langle L,M \rangle, \nonumber \\
&& \headWd \hspace{3.5pt} s(X',Y') \label{wrong}
\end{eqnarray}
Next, we could fold~(\ref{wrong})
using the following definitions:
\setlength{\arraycolsep}{2pt}
\begin{eqnarray*}
\naiveh_0( X, X' ) & \leftrightarrow &
\exists A \, \exists N \, ( X=\langle [A|N] \rangle \ \& \
                                      X'=\langle N \rangle ) \\
\naiveh_1( Y', Y ) & \leftrightarrow &
\exists A \, \exists L \, \exists M \, ( Y'=\langle L,M \rangle \ \& \
                                      Y=\langle [A|L],M \rangle )
\end{eqnarray*}
\setlength{\arraycolsep}{0pt}%
In the case of $\naiveh_0$, this folding operation would replace the
subgoals enclosed in rectangles by the definiendum $\naiveh_0( X, X' )$.
However, as observed 
in~\cite{tst84}, 
in general it is
incorrect to fold a clause such as~(\ref{wrong}) 
using a definition such as that of $\naiveh_0$, 
where a variable like $A$: 
(a)~appears in atoms replaced by the definiendum (i.e.\ 
$X=\langle [A|N] \rangle$),
(b)~appears in atoms not replaced by the definiendum (i.e.\ 
$Y=\langle [A|L],M \rangle$), and
(c)~does not unify with any variable appearing in the definiendum.
To see this incorrectness, fold and subsequently unfold~(\ref{wrong})
with $\naiveh_0$, and a generalisation of~(\ref{wrong}) is obtained.

Note that unlike $A$ in~(\ref{s2}), $N$ (which does not cause the
incorrectness of Tamaki and Sato) occurs only in $t_0$ and $t_0'$.
This suggests that a minimal strengthening of the syntactic
conditions defining moded clauses so as to avoid the situation
above would be requiring that all variables of $t_i'$
occur in $t_i$ (for $i=0,\ldots,n$).
Hence, we define a clause in {\em prechain form} as:
\begin{eqnarray*}
&& \hat{p}(s_0,s_n') \leftarrow
   \hat{q}_1(s_0',s_1),\ \hat{q}_2(s_1',s_2),\ \ldots,\
   \hat{q}_n(s_{n-1}',s_n) \hspace{1cm} n\geq 0 \label{relay}
\end{eqnarray*}
where:
\begin{enumerate}
\item $\var(s_i') \subseteq \var(s_i)$ for $i=0,\ldots,n$; and
\item $\var(s_i) \cap \var(s_j) = \emptyset$,
for $i,j=0,\ldots,n$ and $i \not = j$.
\end{enumerate}

From a clause in prechain form it is possible to arrive at chain form
by first applying
predicate substitutivity 
and then folding
with the completed definitions of predicates of the form:
\begin{eqnarray*}
&& h_i( s_i, s_i' ) \leftarrow
\end{eqnarray*}
Such foldings
satisfy also Gardner and Shepherdson's stronger condition~\cite{gsh92}
for folding.

We face now the problem of converting a moded clause into prechain
form.
Recall first the pass-on variables 
of the deterministic, exhaustive-traversal methods~\cite{ud87,tmk87},
i.e.\ variables that receive a substitution before a subgoal $B$ is
selected and occur in a subgoal selected after $B$ has succeeded.
Observe next that prechain form clauses 
lack pass-on variables.
Finally, note that
the fact that these methods use a stack to hide pass-on variables,
suggests that {\em we also use a stack to achieve prechain form}.

Since both chain form and prechain form lack
pass-on variables, we could have chosen prechain 
form for the object programs of our deterministic metainterpreter.
The resulting metainterpreter, however, would not have been as
concise. 

In converting a moded clause into prechain form,
we will often apply the following sequence of unfolding steps, which we
group in a lemma.
\begin{lemma}[Equation introduction]
\label{eq:int}
Let $C$ be a definite clause having an occurrence of a variable $X$.
Then the clause obtained by replacing that occurrence of $X$ by
$X'$ and adding 
the equation $X=X'$ 
is logically implied by $C$ and
the standard equality theory for $C$.
\end{lemma}

\begin{proof}
First, we resolve $C$ with predicate substitutivity as follows. 
If the occurrence of $X$ is in the head of $C$, then we select the
subgoal of predicate substitutivity which is not an equation.
Next, we apply symmetry to all equations of the resolvent.
The resulting clause has the form:
\[
p(Y_1,\ldots,Y_d) \leftarrow t_1=Y_1,\ \ldots,\ t_d=Y_d,\
A_1,\ \ldots,\ A_n
\]
If, on the other hand, the occurrence of $X$ is in a subgoal $A_i$ of $C$,
then we select such a subgoal.
The resulting clause has the form:
\[
A_0 \leftarrow A_1,\ \ldots,\ A_{i-1},\ 
t_1=Y_1,\ \ldots,\ t_d=Y_d,\ p(Y_1,\ldots,Y_d),\ A_{i+1},\ \ldots,\ A_n
\]
In either case, the occurrence of $X$ is in some equation $t_j=Y_j$.

Next, we apply function substitutivity to $t_j=Y_j$ as many times as it
is necessary to make such an occurrence appear at the top level of
an equation $X=X'$.

Finally, we apply reflexivity
to all equations except $X=X'$,
thus disposing of all unwanted equations.
The resulting clause has the claimed form. 
Hence the lemma holds.
\end{proof}

Often, we will use equation introduction followed by symmetry, to which
we will also refer as equation introduction.

\begin{example}
\label{split:ex}
Let us convert~{\rm (\ref{s2})} first into prechain form and
then into chain form.
To achieve prechain form, we will use an auxiliary stack, which we
introduce in the following predicate:
\begin{eqnarray*}
&& \hs( \langle \St, X \rangle, \langle \St, Y, Z \rangle )
\leftarrow s(\langle X \rangle, \langle Y,Z \rangle )
\end{eqnarray*}
We first apply equation introduction to the definition of $\hats$:
\begin{eqnarray}
&& \hats(\langle \St,X \rangle, \langle \St',Y,Z \rangle )
\leftarrow \St=\St',\ 
\underline{s(\langle X \rangle, \langle Y,Z \rangle )}
\label{h}
\end{eqnarray}
Next, we apply equation introduction to~{\rm (\ref{s2})}.
\begin{eqnarray}
&& \underline{s( \langle [A|N] \rangle, \langle [A'|L],M \rangle )}
\leftarrow A=A',\ s(\langle N \rangle,\langle L,M \rangle ) \label{b}
\end{eqnarray}
Subsequently, we resolve~{\rm (\ref{h})} and~{\rm (\ref{b})} unifying
the two underlined atoms.
\begin{eqnarray*}
&& \head{\hats( \langle \St,[A|N] \rangle, \langle \St',[A'|L],M \rangle )}
\leftarrow \fbox{$\St=\St'$},\ \fbox{$A=A'$},\\
&& \headWd \hspace{3.5pt} s(\langle N \rangle,\langle L,M \rangle )
\end{eqnarray*}
Next, we fold using function substitutivity.
\begin{eqnarray*}
&& \head{\hats( \langle \St,[A|N] \rangle, \langle \St',[A'|L],M \rangle )}
\leftarrow \fbox{$[A|\St] = [A'|\St']$},\\
&& \headWd \fbox{$s(\langle N \rangle,\langle L,M \rangle)$}
\end{eqnarray*}
Finally, we fold using~(\ref{h}) and systematically rename
variables.
\begin{eqnarray*}
&& \hats( \langle \St_0, [A_0|N_0] \rangle, 
     \langle \St_1, [A_1|L_1], M_1 \rangle ) \leftarrow
\hats( \langle [A_0|\St_0], N_0 \rangle, 
                     \langle [A_1|\St_1], L_1, M_1 \rangle )
\end{eqnarray*}

Once we have prechain form, we can apply predicate substitutivity and
symmetry, and
then fold w.r.t.\ the completed definitions of:
\begin{eqnarray*}
&& h_0( \langle \St_0, [A_0|N_0] \rangle, 
        \langle [A_0|\St_0], N_0 \rangle ) \leftarrow \\
&& h_1( \langle [A_1|\St_1], L_1, M_1 \rangle, 
        \langle \St_1, [A_1|L_1], M_1 \rangle ) \leftarrow
\end{eqnarray*}
arriving at:
\begin{eqnarray}
&& \hats(X_0,X_3) \leftarrow h_0(X_0,X_1),\ \hats(X_1,X_2),\
                            h_1(X_2,X_3) \label{split:ch}
\end{eqnarray}
{\it End of example}
\end{example}

In~\cite{rsn96,rpr96} we required that each variable occurring in
$t_i'$ occurs only once in $t_i'$.
Note that if there is more than one occurrence of a variable
in $t_1'$, like in:
\begin{eqnarray*}
&& r(\langle [A|N] \rangle, \langle [A|L],M,\fbox{$A$} \rangle) \leftarrow 
   s(\langle N \rangle, \langle L,M \rangle)
\end{eqnarray*}
thus violating 
such a condition,
we would have an extra equation after applying equation introduction twice:
\begin{eqnarray*}
&& r( \langle [A|N] \rangle, \langle [A'|L],M,A'' \rangle )
\leftarrow A=A',\ \fbox{$A=A''$},\ s(\langle N \rangle,\langle L,M \rangle )
\end{eqnarray*}
Note that we do not wish to eliminate $A=A''$ with reflexivity, since
we would obtain a clause with a variable occurring both in the input
and the output of the head, preventing us from folding the $h$'s.
However, equations such as this one can be eliminated by {\em factoring} all
equations with the same left-hand side.
Hence we have withdrawn
such a condition.

This derivation suggests a proof of a theorem relating a moded clause
with its chain form.

\begin{theorem}
\label{trans}
Let $C$ be a moded clause:
\begin{eqnarray*}
&& p(t_0,t_n') \leftarrow q_1(t_0',t_1),\ q_2(t_1',t_2),\ \ldots,\ 
q_n(t_{n-1}',t_n) 
\hspace{1.95cm} n \geq 0
\end{eqnarray*}
and let
\begin{eqnarray*}
&& \Pi_j =
\big( \Var(t_0) \cup \cdots \cup \Var(t_{j-1}) \big)
\cap
\big( \Var(t_j') \cup \cdots \cup \Var(t_n') \big)
\hspace{0.65cm}
j=0,\ldots,n+1
\end{eqnarray*}
Then the clause $\hat{C}$:
\begin{eqnarray*}
&& \hat{p}(U_0,U_{n}') \leftarrow
h_0(U_0,U_0'),\ \hat{q}_1(U_0',U_1),\ 
h_1(U_1,U_1'),\ \hat{q}_2(U_1',U_2),\ \ldots,\ \\
&& 
\hspace{7.7cm}
\hat{q}_n(U_{n-1}',U_{n}),\ h_n(U_{n},U_{n}')
\end{eqnarray*}
is logically implied by $C$, the standard equality theory for $C$, the
``iff'' version of the function substitutivity axiom for the
list-constructor function symbol:
\begin{eqnarray*}
&& {[}X|Y{]}=[X'|Y'] \leftrightarrow X=X' \; \& \; Y=Y' 
\end{eqnarray*}
and the completed definitions of:
\begin{eqnarray*}
&& \hat{p}(\langle \St|X\rangle,\langle \St|Y\rangle) \leftarrow
   p(X,Y) \\
&& \hat{q}_j(\langle \St|X\rangle,\langle \St|Y\rangle) \leftarrow
   q_j(X,Y) \hspace{1.6cm} j=1,\ldots,n \\
&& h_j(\langle \Sigma_j|t_j\rangle,\langle \Sigma_{j+1}|t_j'\rangle) 
   \leftarrow \hspace{2.64cm} j=0,\ldots,n 
\end{eqnarray*}
where $\Sigma_j$ is any list of the form
$[X_{1,j},\ldots,X_{{d_j},j}|\St]$,
such that
$\{X_{1,j},\ldots,X_{{d_j},j}\} = \Pi_j$,
if $\Pi_j \not = \emptyset$, and
$\Sigma_j$ is $\St$ if $\Pi_j = \emptyset$,
for $j=0,\ldots,n+1$, and the $h_j$'s are predicate symbols not
occurring in $C$.
\end{theorem}

\begin{proof}
First, we apply equation introduction to the definitions of $\hat{p}$
and the $\hat{q}_i$'s.
\begin{eqnarray}
&& \hat{p}(\langle \St|X\rangle,\langle \St'|Y\rangle) \leftarrow
   \St=\St',\ p(X,Y) \label{new:p} \\
&& \hat{q}_j(\langle \St|X\rangle,\langle \St'|Y\rangle) \leftarrow
   \St=\St',\ q_j(X,Y) \hspace{1.3cm} j=1,\ldots,n \label{new:q}
\end{eqnarray}

Let $\sigma_i$, $i=0,\ldots,n$, be 
renaming substitutions~\cite[p.\ 22]{lld87} for $C$ such that:
\begin{eqnarray*}
&& \sigma_i \igualdef \{X/Z : X \in \var(C) \ \& \ 
              Z \not\in \big( \var(C) \cup \bigcup_{j \not = i}
              \rhs(\sigma_j) \big) \}
\end{eqnarray*}
where $\rhs(\sigma_j) = \{Z_1,\ldots,Z_k\}$ 
if $\sigma_j = \{X_1/Z_1,\ldots,X_k/Z_k\}$.

Since the variables in $\rhs(\sigma_i)$ do not occur in $C$ or in any
other $\sigma_j$ ($i \not = j$), 
the application of
$\sigma_i$ to a variable $X$ renames $X$ uniquely. 
Hence, we can think of such an application as the addition of the
subscript $i$ to $X$.

Next, we apply equation introduction to $C$ and rename variables,
obtaining:
\begin{eqnarray*}
&& p(t_0\sigma_0,t_n'\sigma_n) \leftarrow
E,\ q_1(t_0'\sigma_0,t_1\sigma_1),\ q_2(t_1'\sigma_1,t_2\sigma_2),\ \ldots,\ 
q_n(t_{n-1}'\sigma_{n-1},t_n\sigma_n)
\end{eqnarray*}
where
\begin{eqnarray*}
&& E = \{X\sigma_i=X\sigma_k : 
                   X \in \big( \var(t_i) \cap \var(t_k') \big) 
                   \ \& \ i < k \}
\end{eqnarray*}

Subsequently, we resolve with~(\ref{new:p}) and rename $\St$ and $\St'$.
\setlength{\arraycolsep}{-5.5pt}
\begin{eqnarray*}
&& \head{\hat{p}(\langle \St_0|t_0\sigma_0\rangle,
                 \langle \St_n|t_n'\sigma_n\rangle)} \leftarrow
\St_0=\St_n,\ E,\\
&& \headWd
q_1(t_0'\sigma_0,t_1\sigma_1),\ q_2(t_1'\sigma_1,t_2\sigma_2),\ \ldots,\ 
q_n(t_{n-1}'\sigma_{n-1},t_n\sigma_n)
\end{eqnarray*}
\setlength{\arraycolsep}{0pt}%

Let $\Pi \igualdef \bigcup_j \Pi_j$ and
$X \in \Pi$ 
(i.e.\ $X$ occurs in some (and only one) $t_i$ and some $t_k'$, for $i<k$).
We now define
\setlength{\arraycolsep}{2pt}
\begin{eqnarray*}
\phi(X) & = & i, \mbox{ where } X \in \var(t_i) \\
\gamma(X) & = & \max\{k : X \in \var(t_k') \}
\end{eqnarray*}
\setlength{\arraycolsep}{0pt}%
and apply transitivity and factoring, 
replacing the equations by (up to variable renaming):
\[
\{ \St_0=\St_1,\ \St_1=\St_2,\ \ldots,\ \St_{n-1}=\St_n \}
   \cup \bigcup_{X\in\Pi} E_X 
\]
where
\setlength{\arraycolsep}{-5pt}
\begin{eqnarray*}
&& E_X = \{ X\sigma_{\phi(X)}=X\sigma_{\phi(X)+1},\ 
            X\sigma_{\phi(X)+1}=X\sigma_{\phi(X)+2},\ 
\ldots,\ X\sigma_{\gamma(X)-1}=X\sigma_{\gamma(X)} \}
\end{eqnarray*}
\setlength{\arraycolsep}{0pt}%

Note that for all $j$ and all $X$:
\begin{eqnarray*}
&& X \in \big(\var(t_i) \cap \var(t_k')\big) 
\mbox{ for some } i \leq j-1
\mbox{ and some } k \geq j \\
&& \mbox{ iff } \\
&& (X\sigma_{j-1}=X\sigma_j) \in E_X
\end{eqnarray*}
Equivalently, for all $j$ and all $X$:
\[
X \in \bigcup_{i \leq (j-1),\ k \geq j}\big(\var(t_i) \cap \var(t_k')\big)
\mbox{ iff } (X\sigma_{j-1}=X\sigma_j) \in E_X
\]
Hence,
\begin{eqnarray*}
&& \Pi_j = \{X : (X\sigma_{j-1}=X\sigma_j) \in E_X \}
\end{eqnarray*}
for $j=0,\ldots,n+1$,
so that we have exactly all equations for constructing the $\Sigma_j$'s.

Next, we repetitively fold using function substitutivity,
and arrive at:
\setlength{\arraycolsep}{-5pt}
\begin{eqnarray*}
&& \head{\hat{p}(\langle \St_0|t_0\theta_0\rangle,
              \langle \St_n|t_n'\theta_n\rangle)} \leftarrow
   \Sigma_1\theta_0=\Sigma_1\theta_1,\ 
   \Sigma_2\theta_1=\Sigma_2\theta_2,\ \ldots,\ 
   \Sigma_n\theta_{n-1}=\Sigma_n\theta_n,\\
&& \headWd 
q_1(t_0'\theta_0,t_1\theta_1),\ q_2(t_1'\theta_1,t_2\theta_2),\ \ldots,\ 
q_n(t_{n-1}'\theta_{n-1},t_n\theta_n)
\end{eqnarray*}
\setlength{\arraycolsep}{0pt}%

Prechain form is now obtained by folding with~(\ref{new:q}):
\begin{eqnarray*}
&& \head{\hat{p}(\langle\St_0|t_0\theta_0\rangle,
           \langle\St_n|t_n'\theta_n\rangle)} \leftarrow
   \hat{q}_1(\langle\Sigma_1\theta_0|t_0'\theta_0\rangle,
             \langle\Sigma_1\theta_1|t_1\theta_1\rangle), \\
&& \headWd
   \hat{q}_2(\langle\Sigma_2\theta_1|t_1'\theta_1\rangle,
             \langle\Sigma_2\theta_2|t_2\theta_2\rangle), \ \ldots,\\
&& \headWd
   \hat{q}_n(\langle\Sigma_n\theta_{n-1}|t_{n-1}'\theta_{n-1}\rangle,
             \langle\Sigma_n\theta_n|t_n\theta_n\rangle)
\end{eqnarray*}
Finally, we apply predicate substitutivity and symmetry, and then
fold with the completed definitions of the $h_i$ predicates.
The resulting clause has the desired form; hence we conclude that the
theorem holds.
\end{proof}

\begin{example}
Let us apply Theorem~\ref{trans} to the 
clause~{\rm (\ref{s2})}, of
Example~\ref{split:ex}. 
\begin{eqnarray*}
&& \underbrace{\rule[-2.5pt]{0pt}{1cm}s}_{\mbox{$p$}}(
        \underbrace{\langle [A|N] \rangle}_{\mbox{$t_0$}}, 
        \underbrace{\langle [A|L],M \rangle}_{\mbox{$t_1'$}}) 
\leftarrow 
   \underbrace{\rule[-2.5pt]{0pt}{1cm}s}_{\mbox{$q_1$}}( 
        \underbrace{\langle N \rangle}_{\mbox{$t_0'$}}, 
        \underbrace{\langle L,M \rangle}_{\mbox{$t_1$}} )
\end{eqnarray*}

\begin{tabular}{rclcl}
$\Pi_0$ & $=$ & 
   $\emptyset \;\;\cap\;\; \big( \Var(t_0') \cup \Var(t_1') \big)$
                            & $=$ & $\emptyset$ \\
$\Pi_1$ & $=$ & \hspace{4pt} $\Var(t_0) \;\;\cap\;\; \Var(t_1')$ \hfill
                            & $=$ & $\{A\}$ \\
$\Pi_2$ & $=$ & $\big( \Var(t_0) \cup \Var(t_1) \big)
   \;\;\cap\;\; \emptyset$ 
                            & $=$ & $\emptyset$ \\
\end{tabular}

\vspace*{10pt}
\noindent
So, $\Sigma_0 = \St$, $\Sigma_1 = [A|\St]$, and $\Sigma_2 = \St$.
The resulting clause in chain form is~{\rm (\ref{split:ch})}.
The definitions of the $h_i$'s are as before, up to variable renaming.

\noindent {\it End of example}
\end{example}

Let $P$ be a moded program.
We define $\hat{P}$ as the program resulting from applying
Theorem~\ref{trans} to every clause in $P$ in such a way that the
$h_i$ predicate symbols of a clause
do not occur in any other clause of $\hat{P}$.

Theorem~\ref{trans} associates a chain program $\hat{P}$ with a
moded program $P$ in such a way that $\hat{P}$ is a logical
consequence of a conservative extension of $P$.
The implication in the other direction also holds.
That $P$ is logically implied by a conservative extension of $\hat{P}$
can be seen by
first resolving each clause in $\hat{P}$ having a head with predicate
symbol $\hat{p}$ with
the only-if part of the definition of $\hat{p}$ and
unfolding the definitions of
the $h_i$'s and the $\hat{q}_i$'s.

Finally, note that the chain program $\hat{P}$ of a moded program $P$
is 
a G-chain program (i.e.\ a program such that in every unit clause
$p(t,t') \leftarrow$, $\var(t') \subseteq \var(t)$, so that all
answers for a subgoal with a ground input
are ground using a leftmost computation rule (cf.\
Sect.~\ref{meta:int:chain})).

\begin{example}
As an example explicitly linking this transformation with
the abcde metainterpreter, we give the object-program
representation of the chain form of the clauses~{\rm (\ref{s1})}
and~{\rm (\ref{s2})}.
\begin{eqnarray*}
&& \Defn( \hats, [\hats_1,\hats_2] ) \leftarrow \\
&& \Defn( h_0, [h_0'] ) \leftarrow \\
&& \Defn( h_1, [h_1'] ) \leftarrow \\
&& \nonunit( \hats_2, [h_0,\hats,h_1] ) \leftarrow \\
&& \unit( \hats_1, \langle \St,L \rangle, 
                 \langle \St,\nil,L\rangle ) \leftarrow \\ 
&& \unit( h_0', \langle \St, [A|N] \rangle, 
        \langle [A|\St], N \rangle ) \leftarrow \\
&& \unit( h_1', \langle [A|\St], L, M \rangle, 
        \langle \St, [A|L], M \rangle ) \leftarrow \\
&& \isunit( \hats_1 ) \leftarrow \hspace{0.9cm}
   \isunit( h_0' ) \leftarrow \hspace{0.9cm}
   \isunit( h_1' ) \leftarrow 
\end{eqnarray*}

\noindent {\it End of example}
\end{example}

\section{Conversion of Definite Programs into Chain Form}
\label{sect:definite}

In this section, we will first give a transformation inspired by the
previous one, converting an arbitrary definite program into chain
form.
Next, we will explain how to couple the abcde metainterpreter
to this ``unmoded'' transformation.

\subsection{Transformation}
Roughly, the {\em moded} transformation takes a clause with
predicates having one input argument place and one output argument place,
disposes of the pass-on variables (the $\Pi_j$'s) by adding a stack, 
and replaces both arguments of each subgoal by variables.

Hence, the first apparent obstacle we find in trying to convert an
arbitrary, {\em unmoded} clause into chain form,
is that arguments of predicates in such a clause 
do not have predetermined input/output roles.
The fact that any argument of a predicate may play the role of either
input or output suggests treating all arguments uniformly.
One way of doing so and yet have binary predicates
could be to replicate the arguments of each predicate
so as to have two copies of each
set of arguments: one copy behaving as a single input (possibly having
variables when the subgoal is selected),
and the other copy behaving as a single output.
Naturally, we now have to give up the groundness property of runtime
terms, which translates to having to use full unification instead of
matching. 

Thus, we can associate with each predicate $p(X_1,\ldots,X_n)$, 
another predicate, defined as:
$\hat{p}(\langle X_1,\ldots,X_n\rangle,\langle X_1,\ldots,X_n\rangle)
\leftarrow p(X_1,\ldots,X_n)$,
which denotes a {\em subset} of the
identity relation of the Herbrand universe.
But we also have to add the stack, so that we have:
$\hat{p}(\langle \St,X_1,\ldots,X_n\rangle,\langle \St,X_1,\ldots,X_n\rangle)
\leftarrow p(X_1,\ldots,X_n)$.

Another decision we have to make is which variables to push onto the stack.
In fact,
we could push
all variables of the clause, but we will
be more economical by pushing only the variables not occurring in all
atoms of the clause.

\begin{example}
\label{app:ex}
Consider the usual $\append$ program:
\begin{eqnarray}
&& a( \nil, L, L ) \leftarrow
      \label{app1} \\ 
&& \underline{a( [A|L], M, [A|N] )} \leftarrow 
   a( L, M, N )
      \label{app2}
\end{eqnarray}
First we write the if-part of the definition of $\ha$:
\begin{eqnarray}
&& \ha( \langle \St,X,Y,Z \rangle,
        \langle \St,X,Y,Z \rangle ) \leftarrow \label{hata} 
   a( X, Y, Z )
\end{eqnarray}
We now apply equation introduction to~{\rm (\ref{hata})}, 
getting:
\begin{eqnarray}
&& \ha( \langle \St,X,Y,Z \rangle, 
        \langle \St',X',Y',Z' \rangle ) \leftarrow \nonumber \\
&& \hspace{3cm} \St=\St',\ X=X',\ Y=Y',\ Z=Z', \nonumber \\
&& \hspace{3cm} a(X',Y',Z') \label{ha}
\end{eqnarray}
which we will need for a later folding application.

Next, we obtain an instance of~{\rm (\ref{hata})}, to which we apply
equation introduction:
\begin{eqnarray}
&& \ha( \langle \St,[A|L],M,[A|N] \rangle, 
        \langle \St',[A'|L'],M',[A'|N'] \rangle ) \leftarrow
        \nonumber \\
&& \hspace{3cm} \St=\St',\ A=A',\ L=L',\ M=M',\ N=N', \nonumber \\
&& \hspace{3cm} \underline{a([A'|L'],M',[A'|N'])} \label{head} 
\end{eqnarray}
Now we resolve~{\rm (\ref{app2})} with~{\rm (\ref{head})} unifying the
two underlined atoms, and get:
\begin{eqnarray*}
&& \ha( \langle \St,[A|L],M,[A|N] \rangle, 
        \langle \St',[A'|L'],M',[A'|N'] \rangle ) \leftarrow \\
&& \hspace{2.85cm} \fbox{$\St=\St'$},\ \fbox{$A=A'$},\ 
                  L=L',\ M=M',\ N=N',\\
&& \hspace{3cm} a( L', M', N' )
\end{eqnarray*}
Subsequently, we fold using function substitutivity.
\begin{eqnarray*}
&& \ha( \langle \St,[A|L],M,[A|N] \rangle,
        \langle \St',[A'|L'],M',[A'|N'] \rangle ) \leftarrow \\
&& \hspace{2.85cm} \fbox{$[A|\St]=[A'|\St']$},\ 
                  \fbox{$L=L'$},\ \fbox{$M=M'$},\ \fbox{$N=N'$},\\
&& \hspace{2.85cm} \fbox{$a(L',M',N')$}
\end{eqnarray*}
Finally, we fold using~{\rm (\ref{ha})} and systematically rename
variables.
\begin{eqnarray*}
&& \ha( \langle \St_0,[A_0|L_0],M_0,[A_0|N_0] \rangle, 
        \langle \St_1,[A_1|L_1],M_1,[A_1|N_1] \rangle ) \leftarrow \\
&& \hspace{3cm}
   \ha( \langle [A_0|\St_0],L_0,M_0,N_0 \rangle, 
        \langle [A_1|\St_1],L_1,M_1,N_1 \rangle )
\end{eqnarray*}

We can now apply predicate substitutivity and symmetry, and then
fold w.r.t.\ the completed definitions of the predicates manipulating
the stack, as in the moded transformation, thus arriving at
chain form.

\noindent {\it End of example}
\end{example}

An arbitrary definite program can be converted into chain form with
the following theorem.

\begin{theorem}
\label{mmtrans}
Let $C$ be a definite clause:
\begin{eqnarray*}
&& p(\tilde{t}_0) \leftarrow q_1(\tilde{t}_1),\ q_2(\tilde{t}_2),\ 
\ldots,\ q_n(\tilde{t}_n)
\hspace{2cm} n \geq 0
\end{eqnarray*}
and let:
\begin{eqnarray*}
&& \Pi = \big( \Var(\tilde{t}_0) \cup \cdots \cup \Var(\tilde{t}_n) \big) 
\setminus 
      \big( \Var(\tilde{t}_0) \cap \cdots \cap \Var(\tilde{t}_n) \big)
\end{eqnarray*}
Then the clause $C'$:
\begin{eqnarray*}
&& \hat{p}(U_0,U_n') \leftarrow 
h_0(U_0,U_0'),\ \hat{q}_1(U_0',U_1),\ 
h_1(U_1,U_1'),\ \hat{q}_2(U_1',U_2),\ \ldots,\ \\
&& \hspace{7.74cm}
\hat{q}_n(U_{n-1}',U_{n}),\ h_n(U_{n},U_{n}')
\end{eqnarray*}
is logically implied by $C$, the standard equality theory for $C$, the
``iff'' version of the function substitutivity axiom for the
list-constructor function symbol:
\begin{eqnarray*}
&& {[}X|Y{]}=[X'|Y'] \leftrightarrow X=X' \; \& \; Y=Y'
\end{eqnarray*}
and the completed definitions of:
\begin{eqnarray*}
&& \hat{p}(\langle \St,X_{1,0},\ldots,X_{r_0,0} \rangle,
        \langle \St,X_{1,0},\ldots,X_{r_0,0} \rangle ) \leftarrow
p(X_{1,0},\ldots,X_{r_0,0}) \\
&& \hat{q}_i(\langle \St,X_{1,i},\ldots,X_{r_i,i} \rangle,
          \langle \St,X_{1,i},\ldots,X_{r_i,i} \rangle ) \leftarrow
q_i(X_{1,i},\ldots,X_{r_i,i})
\hspace{0.38cm} i=1,\ldots,n \\
&& h_0(\langle \St|\tilde{t}_0\rangle,
       \langle \Sigma|\tilde{t}_1\rangle) \leftarrow
\hspace{1.52cm} (i=0) \\
&& h_i(\langle \Sigma|\tilde{t}_i\rangle,
       \langle \Sigma|\tilde{t}_{i+1}\rangle) \leftarrow
\hspace{1.46cm} i = 1,\ldots,n-1 \\
&& h_n(\langle \Sigma|\tilde{t}_n\rangle,
       \langle \St|\tilde{t}_0\rangle) \leftarrow
\hspace{1.43cm} (i=n)
\end{eqnarray*}
where
$\Sigma$ is any list of the form
$[X_{1},\ldots,X_{d}|\St]$,
such that
$\{X_{1},\ldots,X_{d}\} = \Pi$,
if $\Pi \not = \emptyset$, and
$\Sigma$ is $\St$ if $\Pi = \emptyset$.
\end{theorem}

\begin{proof}
First, we use equation introduction on the definitions of 
the $\hat{q}_i$'s.
\begin{eqnarray}
&& \head{\hat{q}_i(\langle \St,X_{1,i},\ldots,X_{r_i,i} \rangle,
        \langle \St',X_{1,i}',\ldots,X_{r_i,i}' \rangle )} \leftarrow
\St=\St', \nonumber \\
&& \headWd X_{1,i}=X_{1,i}',\ldots,X_{r_i,i}=X_{r_i,i}', \nonumber \\
&& \headWd q_i(X_{1,i}',\ldots,X_{r_i,i}') \nonumber \\
&& \headWd \hspace{1cm} i=1,\ldots,n \label{new:qs}
\end{eqnarray}

Next, we apply equation introduction to $C$
in such a way that no two $\tilde{t}_i$'s
have variables in common, except for $\tilde{t}_0$ and $\tilde{t}_n$.
We obtain:
\begin{eqnarray}
&& p(\tilde{t}_0\sigma_n) \leftarrow E,\  
q_1(\tilde{t}_1\sigma_1),\ q_2(\tilde{t}_2\sigma_2),\ 
\ldots,\ q_n(\tilde{t}_n\sigma_n) \label{mm:body}
\end{eqnarray}
where the $\sigma_j$'s are as in Theorem~\ref{trans} and
\begin{eqnarray*}
       E & = & \{X\sigma_j=X\sigma_k : 
            X \in \big( \var(\tilde{t}_j) \cap \var(\tilde{t}_k) \big)
                   \ \& \ 0 < j < k < n \} \\
    & \union & \{X\sigma_j=X\sigma_n : X \in \Big( \var(\tilde{t}_j) \cap 
                     \big( \var(\tilde{t}_0)\cup\var(\tilde{t}_n) \big) \Big)
                            \ \& \ 0 < j < n \}
\end{eqnarray*}

Now, we apply $\theta$ to the if-part of the
definition of $\hat{p}$, where $\theta$
is the mgu of $p(X_{1,0},\ldots,X_{r_0,0})$ and $p(\tilde{t}_0)$ and
use equation introduction in the resulting instance, 
getting (up to variable renaming):
\begin{eqnarray}
&& \hat{p}(\langle \St_0|\tilde{t}_0\sigma_0 \rangle,
           \langle \St_n|\tilde{t}_0\sigma_n\rangle ) \leftarrow
\St_0=\St_n,\ F,\ p(\tilde{t}_0\sigma_n) \label{mm:head}
\end{eqnarray}
where
\begin{eqnarray*}
&& F = \{X\sigma_0=X\sigma_n : X \in \var(\tilde{t}_0)\}
\end{eqnarray*}

Subsequently, we resolve~(\ref{mm:body}) with~(\ref{mm:head}):
\begin{eqnarray*}
&& \head{\hat{p}(\langle \St_0|\tilde{t}_0\sigma_0 \rangle,
                 \langle \St_n|\tilde{t}_0\sigma_n \rangle )} \leftarrow
\St_0=\St_n,\ F,\ E,\\
&& \headWd q_1(\tilde{t}_1\sigma_1),\ q_2(\tilde{t}_2\sigma_2),\ 
\ldots,\ q_n(\tilde{t}_n\sigma_n)
\end{eqnarray*}

Before folding, we add the following equations, recalling that subgoal
addition preserves soundness:
\begin{eqnarray*}
&& \{X\sigma_0=X\sigma_n : X \in \big(\Pi \setminus \var(\tilde{t}_0)\big)\}
\end{eqnarray*}
(Such equations, after applying transitivity,
will enable us to have the same $\Sigma$ in each
subgoal of the resulting clause in prechain form.)

Now we add the equations:
\begin{eqnarray*}
&& \{X\sigma_{i-1}=X\sigma_i : X \in \var(\tilde{t}_i) \} \hspace{2cm}
0 < i \leq n
\end{eqnarray*}
(Such equations will enable us to fold w.r.t.~(\ref{new:qs}).)

We now apply transitivity and factoring
in such a way that the equations are replaced by (up to variable renaming):
\begin{eqnarray*}
&& \{\St_0=\St_1,\ \St_1=\St_2,\ \ldots,\ \St_{n-1}=\St_n\}
   \cup \bigcup_{0 < i \leq n} G_i \cup \bigcup_{0 < i \leq n} H_i
\end{eqnarray*}
where
\setlength{\arraycolsep}{2pt}
\begin{eqnarray*}
G_i & = & \{ X\sigma_{i-1}=X\sigma_i : X \in \Pi \} \\
H_i & = & \{ X\sigma_{i-1}=X\sigma_i : X \in \var(\tilde{t}_i) \}
\end{eqnarray*}
\setlength{\arraycolsep}{0pt}%

Next, we repetitively fold using function substitutivity
and arrive at:
\begin{eqnarray*}
&& \head{\hat{p}(\langle \St_0|\tilde{t}_0\sigma_0 \rangle,
                 \langle \St_n|\tilde{t}_0\sigma_n \rangle )} \leftarrow
\Sigma\sigma_0=\Sigma\sigma_1,\ \hspace{1.1pt}
\Sigma\sigma_1=\Sigma\sigma_2,\ \hspace{1.1pt} \ldots,\ 
\Sigma\sigma_{n-1}=\Sigma\sigma_n, \\
&& \headWd \tilde{t}_1\sigma_0 \cong \tilde{t}_1\sigma_1, \ 
           \tilde{t}_2\sigma_1 \cong \tilde{t}_2\sigma_2, \ \ldots,\ 
           \tilde{t}_n\sigma_{n-1} \cong \tilde{t}_n\sigma_n, \\
&& \headWd q_1(\tilde{t}_1\sigma_1),\ \hspace{14.5pt} 
           q_2(\tilde{t}_2\sigma_2),\ \hspace{14.5pt} \ldots,\ 
           q_n(\tilde{t}_n\sigma_n)
\end{eqnarray*}
where 
\setlength{\arraycolsep}{2pt}
\begin{eqnarray*}
\tilde{r} \cong \tilde{s} & = & \{r^i=s^i : 
      r^i \mbox{ is the $i^{\mbox{\scriptsize th}}$ term of $\tilde{r}$ and }
      s^i \mbox{ is the $i^{\mbox{\scriptsize th}}$ term of $\tilde{s}$} \}
\end{eqnarray*}
\setlength{\arraycolsep}{0pt}%

Prechain form is now obtained by folding with~(\ref{new:qs}):
\begin{eqnarray*}
&& \head{\hat{p}(\langle \St_0|\tilde{t}_0\sigma_0 \rangle,
             \langle \St_n|\tilde{t}_0\sigma_n\rangle )} \leftarrow
   \hat{q}_1(\langle\Sigma\sigma_0|\tilde{t}_1\sigma_0\rangle,
             \langle\Sigma\sigma_1|\tilde{t}_1\sigma_1\rangle),\\
&& \headWd \hat{q}_2(\langle\Sigma\sigma_1|\tilde{t}_2\sigma_1\rangle,
             \langle\Sigma\sigma_2|\tilde{t}_2\sigma_2\rangle),\ \ldots,\\
&& \headWd \hat{q}_n(\langle\Sigma\sigma_{n-1}|\tilde{t}_n\sigma_{n-1}\rangle,
             \langle\Sigma\sigma_n|\tilde{t}_n\sigma_n\rangle)
\end{eqnarray*}
Finally, we apply predicate substitutivity and symmetry, and then
fold with the completed definitions of the $h_i$ predicates.
The resulting clause has the desired form; hence we conclude that the
theorem holds.
\end{proof}

As in the moded transformation, 
that $P$ is logically implied by a conservative extension of $\hat{P}$
can be seen by 
unfolding the definitions of
the $h_i$'s and the $\hat{q}_i$'s.

\subsection{A deterministic metainterpreter for arbitrary chain programs}

In Sect.~\ref{meta:int:chain} we wrote a deterministic metainterpreter
assuming that the leftmost input in every goal of the LD
tree~\cite{apt97} (i.e.\ an SLD tree with a leftmost computation rule)
was ground.
We will now
modify such a metainterpreter so that it
is also correct for LD trees that do not necessarily have this
groundness property.

It is possible to handle terms with variables by generalising matching
to full unification.
A well-known metainterpreter explicitly using unification is Bowen and
Kowalski's $\demo$ metainterpreter~\cite{bkw82}.
In our case, unification for clauses of the form~(\ref{chain:cl})
reduces to argument passing so that we need only incorporate it to
the clauses of the form~(\ref{prim:cl}).
Consider for instance our previous definition of $e'$, which included
the clause: 
\begin{eqnarray*}
&& e'(X,\Pj,[Y]) \leftarrow \unit(\Pj,X,Y) 
\end{eqnarray*}
Following the $\demo$ metainterpreter, we would replace this clause by:
\begin{eqnarray}
&& \head{e'(X,\Pj,[Y])} \leftarrow \unit(\Pj,X',Y'), \nonumber \\ 
&& \headWd \Rename(f(X',Y'),X,f(X'',Y'')), \nonumber \\
&& \headWd \match(X,X'',\Sub), \nonumber \\
&& \headWd \apply(Y'',\Sub,Y) \label{mm:e}
\end{eqnarray}
where 
\begin{enumerate}
\item $\Rename(Z,X,Z')$ holds when $Z'$ is the result
of renaming the variables in $Z$ so that they are distinct from the
variables in $X$,
\item $\match(X,X'',\Sub)$ holds when $\Sub$ is the mgu of $X$ and
$X''$, and
\item $\apply(Y'',\Sub,Y)$ holds when $Y$ is the result of applying
$\Sub$ to $Y''$.
\end{enumerate}

Prolog provides a way to approximate this effect through the
extralogical $\Copy$ ``predicate'':
\begin{eqnarray}
&& e'(X,\Pj,[Y]) \leftarrow \Copy(X,\Xtheta),\ \unit(\Pj,\Xtheta,Y)
\label{mm:e:copy}
\end{eqnarray}

Note that our addition of unification to the abcde metainterpreter 
occurs at a single point, unlike the
unifiers of the metainterpreter in Fig.~\ref{clgr}, which are
pervasive (cf.\ the variables with a $\theta$ in their name).

\section{Extending Existing Committed-Choice Traversal Methods to Arbitrary
Definite Programs}
\label{reconstr}

This section deals first with a reconstruction and then 
with an extension of the existing stream-based~\cite{tmk87}
and continuation-based~\cite{ud87}
deterministic, exhaustive-traversal methods.
An objective of these methods is that of executing
\Or-parallel programs in committed-choice processors (which are 
\And\ parallel).

The existing versions of such methods are restricted to moded programs.
Hence, our reconstruction uses our transformation of moded programs
into G-chain form.
The extension modifies such methods so as to make them applicable to
arbitrary definite programs essentially
by replacing the moded transformation of
Sect.~\ref{sect:moded} by the definite transformation of
Sect.~\ref{sect:definite}. 

We fall short of proposing practical methods because we do not
eliminate the layer of interpretation.
One way of eliminating such a layer would be to feed the
metainterpreter and the object program to a general-purpose partial
evaluator such as Mixtus~\cite{shl93}.
However,
the resulting residual program may be
enormous.
Another possibility would be to compile away the layer of
interpretation ``by hand,'' but we have not done so in the present
work.

\subsection{Reconstruction}
The derivations of both methods start from the abcde metainterpreter.
For brevity, we will omit detailed derivations, and will only
indicate how such derivations could be obtained.
Also, instead of using difference lists as the original methods do, 
we will employ ordinary lists for clarity.

\subsubsection{A chain-program reconstruction of the stream-based method}
As observed by Tamaki, programs originally having some degree
of \And\ parallelism may lose such a parallelism if we only capture their
\Or\ parallelism.
He thus treats clauses with \And\ parallelism in a special way.
For simplicity we will not be concerned with such a special treatment
here, 
and will concentrate on the main
component of this method,
that converts \Or\ parallelism into \And\ parallelism.

We can obtain programs produced by the
stream-based method if we
unfold~(\ref{abcd:a}) using~(\ref{abcd:a1}):
\begin{eqnarray}
&& \head{a( [X|\Xs], [Q|\Qs], \YsZs )} \leftarrow
        b'( X, Q, \Ys ),\ 
        a( \Ys, \Qs, \Ys' ), \nonumber \\
&& \headWd a( \Xs, [Q|\Qs], \Zs ), \ 
                   \append( \Ys', \Zs, \YsZs ) \label{unf:a}
\end{eqnarray}

Let us consider first how a chain program is transformed by the
stream-based method.
Perhaps the most interesting clause in the program produced by
this method is a clause associated with each subgoal
$q_{i+1}(X_i,X_{i+1})$, which is of the form:
\begin{eqnarray}
&& \head{k_i([X|\Xs],\YsZs)} \leftarrow \allq_{i+1}(X,\Ys),\
k_{i+1}(\Ys,\Ys'), \nonumber \\ 
&& \headWd k_i(\Xs,\Zs),\ \append( \Ys', \Zs, \YsZs ) \label{k}
\end{eqnarray}
where $\allq_{i+1}(\x,\ys)$ is intended to hold when $\ys$ is the set
of answers to \linebreak $\leftarrow q_{i+1}(\x,Y)$.
Observe first that $\allq_{i+1}(\x,\ys)$ has the same intended meaning
as
$b'(\x,q_{i+1},\ys)$.
Next, it is easy to obtain~(\ref{k}) from~(\ref{unf:a}) by identifying
the $k_i$ predicate with the $a$ predicate.
The rest of the clauses resulting from transforming a chain program
are readily obtainable from the abcde metainterpreter.

Consider now an arbitrary moded program.
The clause corresponding to~(\ref{k})
in this case is:
\begin{eqnarray}
&& \head{k_i(\pi_i,[t_i|\Xs],\YsZs)} \leftarrow \allq_{i+1}(t_i',\Ys),\
   k_{i+1}(\pi_{i+1},\Ys,\Ys'),\nonumber \\
&& \headWd k_i(\pi_i,\Xs,\Zs), \ 
                  \append( \Ys', \Zs, \YsZs ) \label{kpi}
\end{eqnarray}
where $\pi_j$ is any term such that $\var(\pi_j)=\Pi_j$, and $\Pi_j$
is as defined in Theorem~\ref{trans}.

Recall now~(\ref{unf:a}) and an object clause transformed by our moded
transformation. 
A subgoal with an $h_i$ predicate symbol, when partially evaluated
with the abcde metainterpreter, results in a subgoal of the
form $b'(X,h_i,\Ys)$, which can be easily unfolded away producing the
following clause, similar to~(\ref{kpi}):
\begin{eqnarray}
&& \head{a( [\langle\Sigma_i|t_i\rangle|\Xs], [Q|\Qs], \YsZs )} \leftarrow
	b'( \langle\Sigma_{i+1}|t_i'\rangle, Q, \Ys ),\ 
	a( \Ys, \Qs, \Ys' ), \nonumber \\
&& \headWd a( \Xs, [Q|\Qs], \Zs ),\
                \append( \Ys', \Zs, \YsZs ) \label{kpi:abcd}
\end{eqnarray}
A difference between~(\ref{kpi}) and~(\ref{kpi:abcd}) 
is that the stream-based method uses
a parameter $\pi_i$ in the $k_i$ predicates for recording the values
of the $\Pi_i$ variables,
whereas~(\ref{kpi:abcd}) keeps the $\Pi_i$
variables as part of each term $\langle\Sigma_i|t_i\rangle$.
Even with this difference, the stream-based compiled program and
the abcde metainterpreter follow the same search strategy.
By using a separate parameter $\pi_i$, however, the stream-based
method is more economical because of exploiting the  
fact that only the $h_i$ predicates may modify the stack: all other
predicates hold for relations with an output stack equal to the input stack.
(To see this, observe the stack
in the definitions of $\hat{p}$, $\hat{q}_i$ in
Theorem~\ref{trans}.)
Hence, we can compute all answers to a goal 
$\leftarrow b'(\langle \Sigma_{i+1}|t_i'\rangle,Q,\Ys)$,
by first computing all answers to $\leftarrow b'(t_i',Q,\Ws)$ and then
affixing $\Sigma_{i+1}$ in front of each such answer, if $Q$ is not an
$h_i$ predicate:
\begin{eqnarray}
\langle\St|t\rangle \comp Q = \St \Insert (t \comp Q)
\label{equal:stack}
\end{eqnarray}
where 
$X \Insert \Xs$ is the list obtained by concatenating all lists
having $X$ affixed in front of every list in $\Xs$:
\setlength{\arraycolsep}{2pt}
\begin{eqnarray*}
X \Insert \nil & = & \nil \\
X \Insert [Y|\Ys] & = & [\hspace{1pt}[X|Y]\hspace{1pt}|X \Insert \Ys] 
\end{eqnarray*}
\setlength{\arraycolsep}{0pt}%

Let us first rewrite the stream-based clause~(\ref{kpi}) as:
\setlength{\arraycolsep}{-5pt}
\begin{eqnarray}
&& \head{a_{\stream}(\Sigma_i,[t_i|\Xs],[Q|\Qs],\YsZs)} \leftarrow 
         b'(t_i',Q,\Ys),\
         a_{\stream}(\Sigma_{i+1},\Ys,\Qs,\Ys'),\nonumber \\
&& \headWd a_{\stream}(\Sigma_i,\Xs,[Q|\Qs],\Zs), \nonumber \\
&& \headWd \append( \Ys', \Zs, \YsZs ) \label{api}
\end{eqnarray}
\setlength{\arraycolsep}{0pt}%
where the predicate 
$a_{\stream}(St, \Xs, \Qs, \YsZs )$ 
is intended to hold iff 
\[
(\St \Insert \Xs) \comp \Qs = \St \Insert \YsZs
\]

Hence,~(\ref{api}) asserts that:
\begin{eqnarray}
&& \Sigma_i \Insert [t_i|\Xs] \comp (Q \comp \Qs) \supseteq
   \big( \Sigma_{i+1} \Insert (t_i' \comp Q) \comp \Qs \big) \concat
   \big( \Sigma_i \Insert \Xs \comp (Q \comp \Qs) \big)
\label{api:1}
\end{eqnarray}
Here, and throughout this section, 
$\Insert$ binds stronger than ``$\comp$'', and
$\concat$ denotes list concatenation.

Let us now consider the abcde metainterpreter.
For clarity, we rename variables in~(\ref{kpi:abcd}):
\setlength{\arraycolsep}{-5pt}
\begin{eqnarray}
&& \head{a( [\langle\Sigma_i|t_i\rangle|\StXs], [Q|\Qs], \StYsZs )} \leftarrow
	b'( \langle\Sigma_{i+1}|t_i'\rangle, Q, \StYs ),\ 
	a( \StYs, \Qs, \StYs' ), \nonumber \\
&& \headWd a( \StXs, [Q|\Qs], \StZs ), \nonumber \\
&& \headWd \append( \StYs', \StZs, \StYsZs ) \label{api:abcd}
\end{eqnarray}
\setlength{\arraycolsep}{0pt}%
where
$a( \StXs, \Qs, \StYsZs )$ holds when $\StXs \comp \Qs = \StYsZs$ and 
the same stack $\Sigma_i$ occurs
in front of every element of $\StXs$ and $\StYsZs$. 
Hence,~(\ref{api:abcd}) asserts that:
\begin{equation}
[\langle\Sigma_i|t_i\rangle|\Sigma_i\Insert\Xs] \comp (Q \comp \Qs) 
   \supseteq
   \big( (\langle \Sigma_{i+1}|t_i'\rangle \comp Q) \comp \Qs \big) \concat
   \big( \Sigma_i\Insert\Xs \comp (Q \comp \Qs) \big) \label{api:abcd:1}
\end{equation}
By the definition of $\Insert$ and~(\ref{equal:stack}), 
we obtain~(\ref{api:1}) from~(\ref{api:abcd:1}).
(Specifically, by the recursive equation in the definition of
$\Insert$ we obtain the left-hand side
and by~(\ref{equal:stack}) we obtain the right-hand side.)

\subsubsection{A chain-program reconstruction of the
continuation-based method} 
Let us now turn our attention to the continuation-based method.
Starting also from the abcde metainterpreter, we use the completed
definitions of:
\begin{eqnarray}
&& c'\hspace{-3pt}\_a(X,\Pjs,\Qs,M) \leftarrow 
   \underline{c'(X,\Pjs,L)},\ a(L,\Qs,M)
\label{c:a} \\
&& d'\hspace{-3pt}\_a(X,\Pj,\Qs,M) \leftarrow 
\underline{d'(X,\Pj,L)},\ a(L,\Qs,M) 
\label{d:a}
\end{eqnarray}
where $\Qs$ acts like a list of ``continuations,''
and the underlined subgoals indicate a forthcoming unfolding application.

First we unfold~(\ref{c:a}) using~(\ref{abcd:c}):
\begin{eqnarray}
&& \head{c'\hspace{-3pt}\_a(X,[\Pj|\Pjs],\Qs,M)} \leftarrow 
   d'(X,\Pj,L_1),\ c'(X,\Pjs,L_2), \nonumber \\
&& \headWd \append(L_1,L_2,L),\ a(L,\Qs,M) \label{c:a1}
\end{eqnarray}
Using now the identity
\begin{eqnarray*}
&& \big( (X \comp P_1) \cup \ldots \cup (X \comp P_m) \big) \comp \Qs =
    (X \comp P_1 \comp \Qs) \cup \ldots \cup (X \comp P_m \comp \Qs)
\end{eqnarray*}
which follows from the right distributivity of composition over union,
we rewrite~(\ref{c:a1}) as:
\begin{eqnarray}
&& \head{c'\hspace{-3pt}\_a(X,[\Pj|\Pjs],\Qs,M)} \leftarrow 
   \fbox{$d'(X,\Pj,L_1),\ a(L_1,\Qs,M_1)$}, \nonumber \\
&& \headWd \fbox{$c'(X,\Pjs,L_2),\ a(L_2,\Qs,M_2)$}, \nonumber \\
&& \headWd \hspace{3.5pt} \append(M_1,M_2,M) \label{c:a2} 
\end{eqnarray}
where each rectangle indicates a forthcoming folding application.
We can now fold~(\ref{c:a2}) using the definitions of the 
$d'\hspace{-3pt}\_a$ and
$c'\hspace{-3pt}\_a$ predicates:
\begin{eqnarray}
&& \head{c'\hspace{-3pt}\_a(X,[\Pj|\Pjs],\Qs,M)} \leftarrow 
   d'\hspace{-3pt}\_a(X,\Pj,\Qs,M_1), \nonumber \\
&& \headWd c'\hspace{-3pt}\_a(X,\Pjs,\Qs,M_2), \nonumber \\
&& \headWd \append(M_1,M_2,M) \label{c:a3}
\end{eqnarray}
arriving at a clause of the continuation-based metainterpreter.

Another interesting clause is obtained by unfolding~(\ref{d:a}) 
using~(\ref{abcd:d}):
\begin{eqnarray*}
&& d'\hspace{-3pt}\_a(X,\Pj,\Qs',M) \leftarrow 
\nonunit(\Pj,\Qs),\ a'(X,\Qs,L),\ a(L,\Qs',M) \label{d:a1} 
\end{eqnarray*}
which we rewrite as:
\begin{eqnarray*}
&& \head{d'\hspace{-3pt}\_a(X,\Pj,\Qs',M)} \leftarrow \nonunit(\Pj,\Qs),\\
&& \headWd \append(\Qs,\Qs',\QsQs'),\ a'(X,\QsQs',M)
\end{eqnarray*}
This step can be justified using the associativity of composition:
\begin{eqnarray*}
&& (X \comp \Qs) \comp \Qs' = X \comp (\Qs \comp \Qs')
\end{eqnarray*}
Figure~\ref{ueda} shows the resulting metainterpreter, where we have
applied an unfolding step using the definition of $e'$.
\begin{figure}[hbt]
\small
\figrule
\begin{eqnarray*}
&& a'(X,\nil,[X]) \leftarrow \\
&& a'(X,[Q|\Qs],\Zs) \leftarrow \Defn(Q,\Pjs),\
c'\hspace{-3pt}\_a(X,\Pjs,\Qs,\Zs) \\ 
&& c'\hspace{-2pt}\_a(X,\nil,\Qs,\nil) \leftarrow \\
&& \head{c'\hspace{-2pt}\_a(X,[\Pj|\Pjs],\Qs,\YsZs)} \leftarrow 
           d'\hspace{-2pt}\_a(X,\Pj,\Qs,\Ys),\ 
           c'\hspace{-2pt}\_a(X,\Pjs,\Qs,\Zs),\\
&& \headWd \append(\Ys,\Zs,\YsZs) \\
&& d'\hspace{-2pt}\_a(X,\Pj,\Qs,\Zs) \leftarrow \isunit(\Pj),\ 
           \unit(\Pj,X,Y),\ 
           a'(Y,\Qs,\Zs) \\
&& d'\hspace{-2pt}\_a(X,\Pj,\Qs,\nil) \leftarrow \isunit(\Pj),\ 
                                          \Not(\unit(\Pj,X,Y)) \\
&& d'\hspace{-2pt}\_a(X,\Pj,\Qs',\Zs) \leftarrow \nonunit(\Pj,\Qs),\
                                    \append(\Qs,\Qs',\QsQs'),\ 
                                    a'(X,\QsQs',\Zs)
\end{eqnarray*}
\caption{\it A continuation-based, deterministic, exhaustive-traversal
metainterpreter.}
\label{ueda}
\figrule
\end{figure}

\subsection{Unmoded versions of the stream-based and the
continuation-based methods}

Having reconstructed the stream- and continuation-based methods
through chain programs, we can now replace the moded transformation by
the definite transformation.
However, as in Sect.~\ref{sect:definite}, we must also
rename variables
when using the $\unit$
predicate with
either~(\ref{mm:e}) or~(\ref{mm:e:copy}).

\section{Other Applications}
\label{other:app}

\subsection{Prolog as the continuation-based metainterpreter together
with the definite transformation}

So far we have designed metainterpreters for
performing
traversals in committed-choice processors.
Observe that just as committed-choice processors have deterministic
bindings, so do standard (deterministic)
imperative languages.
This suggests the possibility of using our 
metainterpreters for describing search strategies in one such 
imperative language.
In particular, we will see how to obtain an imperative
implementation of 
Prolog's search strategy
by slightly modifying the continuation-based metainterpreter of
Sect.~\ref{reconstr} (Fig.~\ref{ueda}).

A difference between our previous metainterpreters and the standard
implementations of Prolog is that whereas we perform exhaustive
traversals, Prolog systems may or may not do so.
However, we can easily modify the continuation-based metainterpreter
so as to ask the user whether or not more answers are requested.

Also, note that Prolog systems do not
usually remember the answers to a query, allowing us to
eliminate the answer list.
\begin{eqnarray*}
&& a'( X, \nil, \halt ) \leftarrow
        \Write( X ),\ 
        \Write( \More ),\ 
        \Read( n ) \\
&& a'( X, \nil, \cont ) \leftarrow \\
&& a'( X, [Q|\Qs], \Halt ) \leftarrow
        \Defn( Q, \Pjs ),\ 
        c'\hspace{-3pt}\_a( X, \Pjs, \Qs, \Halt ) \\ 
&& \ \\
&& c'\hspace{-3pt}\_a( X, \nil, \Qs, \cont ) \leftarrow \\
&& \head{c'\hspace{-3pt}\_a( X, [\Pj|\Pjs], \Qs, \Halt' )} \leftarrow
        d'\hspace{-3pt}\_a( X, \Pj, \Qs, \Halt ),\\
&& \headWd
        \goon( X, \Pjs, \Qs, \Halt, \Halt' ) \\
&& \ \\
&& d'\hspace{-3pt}\_a( X, \Pj, \Qs, \Halt ) \leftarrow
        \isunit( \Pj ),\ 
        \unit( \Pj, X, Y ),\ 
	a'( Y, \Qs, \Halt ) \\
&& d'\hspace{-3pt}\_a( X, \Pj, \Qs, \cont ) \leftarrow
	\isunit( \Pj ),\ 
	\Not( \unit( \Pj, X, Y ) ) \\
&& \head{d'\hspace{-3pt}\_a( X, \Pj, \Qs', \Halt )} \leftarrow
	\nonunit( \Pj, \Qs ),\\
&& \headWd
	\append( \Qs, \Qs', \QsQs' ),\ 
	a'( X, \QsQs', \Halt ) \\
&& \ \\
&& \goon( X, \Pjs, \Qs, \halt, \halt ) \leftarrow \\ 
&& \goon( X, \Pjs, \Qs, \cont, \Halt ) \leftarrow
        c'\hspace{-3pt}\_a( X, \Pjs, \Qs, \Halt )
\end{eqnarray*}

This metainterpreter is meant for constructing ground terms.
To obtain a true (pure) Prolog system, handling terms with variables,
we would have to include variable renaming and
unification in a manner similar to that of
either~(\ref{mm:e}) or~(\ref{mm:e:copy}).

Deterministic metainterpreters for arbitrary definite programs can be
written directly, without using our transformations, into chain form.
(An example is the metainterpreter in Fig.~\ref{clgr}.)
The programmer, however, has to be aware 
of the set of unifiers of
the children of a
node in the derivation tree.
By contrast, in our approach the programmer can write a
metainterpreter without considering such unifiers, except when using
the $\unit$ 
predicate, in which case the object clause has no body,
simplifying the treatment of unifiers.

\subsection{A bounded-resource metainterpreter}
As a final application, we exhibit a bounded-resource metainterpreter.
As usual, we will give a metainterpreter for chain programs.
Our transformations of moded and definite programs into chain form
make this metainterpreter applicable to programs that do not
necessarily have chain form.

The next variant of the abcde metainterpreter constructs at most one
proof, and has an extra argument to indicate the amount of 
resources needed to construct such a proof.

\begin{eqnarray*}
&& a'( X, \nil, \ans(X), 0 ) \leftarrow \\
&& a'( X, [Q|\Qs], Z, R+1 ) \leftarrow
        \Defn( Q, \Pjs ),\ 
        c'\hspace{-3pt}\_a( X, \Pjs, \Qs, Z, R ) \\ 
&& \ \\
&& c'\hspace{-3pt}\_a( X, \nil, \Qs, \noans, 0 ) \leftarrow \\
&& \head{c'\hspace{-3pt}\_a( X, [\Pj|\Pjs], \Qs, Z, R+S )} \leftarrow
        d'\hspace{-3pt}\_a( X, \Pj, \Qs, Y, R ),\\
&& \headWd
        \goon( X, \Pjs, \Qs, Y, Z, S ) \\
&& \ \\
&& d'\hspace{-3pt}\_a( X, \Pj, \Qs, Z, R ) \leftarrow
        \isunit( \Pj ),\ 
        \unit( \Pj, X, Y ),\ 
	a'( Y, \Qs, Z, R ) \\
&& d'\hspace{-3pt}\_a( X, \Pj, \Qs, \noans, 0 ) \leftarrow
	\isunit( \Pj ),\ 
	\Not( \unit( \Pj, X, Y ) ) \\
&& \head{d'\hspace{-3pt}\_a( X, \Pj, \Qs', Z, R )} \leftarrow
	\nonunit( \Pj, \Qs ),\\
&& \headWd
	\append( \Qs, \Qs', \QsQs' ),\ 
	a'( X, \QsQs', Z, R ) \\
&& \ \\
&& \goon( X, \Pjs, \Qs, \ans(Y), \ans(Y), 0 ) \leftarrow \\ 
&& \goon( X, \Pjs, \Qs, \noans, Z, S ) \leftarrow  
        c'\hspace{-3pt}\_a( X, \Pjs, \Qs, Z, S )
\end{eqnarray*}
This style of writing bounded-resource metainterpreters 
may be viewed as an alternative to that appearing in~\cite{kwl95,ksd96}.

\section{Concluding Remarks}
\label{concl}

\subsection{Contributions}
Some applications of metainterpreters have been neglected, perhaps
because of being based on convoluted definitions.
By comparison with the demo predicate,
deterministic metainterpreters, for example, result especially
elaborate, since the programmer must consider the set of unifiers
of the children of a node in the derivation tree.
Thus, metainterpreters for 
(1)~converting \Or\ parallelism into \And\ parallelism~\cite{ud87,tmk87},
(2)~describing search-strategies in logic-based, state-oriented
languages~\cite{cvn81},
and 
(3)~simulating bounded-resource reasoning~\cite{kwl95,ksd96}, 
have not received due attention. 

Compilation methods converting 
\Or\ parallelism into \And\ parallelism
have been developed first by Ueda~\cite{ud87} and then by
Tamaki~\cite{tmk87}.
By studying these methods, we have identified {\em chain} programs as
important for exhibiting the essence of such techniques.
If we use chain programs as a stepping stone, then the methods
of~\cite{ud87,tmk87} can be viewed as comprising two parts:
\begin{list}
{\alph{contador}.}{\usecounter{contador}}
\item conversion of a {\em moded} program into chain form, and
\item application of partial deduction to a deterministic
metainterpreter for chain programs. 
\end{list}

Our contribution to part (a) consisted first in having
extracted from~\cite{ud87,tmk87} the implicit transformation
that converts a moded program into an equivalent chain form.
Next, by using a generalisation of this transformation, we
have given another, ``unmoded'' transformation, that converts {\em arbitrary}
definite programs into chain form.

To part (b) we contributed by
showing how to write 
deterministic metainterpreters for chain programs.
One such metainterpreter
served us first to reconstruct and then to extend to arbitrary
(unmoded) definite programs the existing methods of~\cite{ud87,tmk87}. 

Finally, we observed that deterministic metainterpreters have
applications other than exhaustive traversals.
We gave a 
metainterpreter that follows Prolog's
search strategy and another one that counts the number of steps in the
search for a refutation (as opposed to the number of steps in the
refutation). 

Our methodology for designing deterministic-traversal methods 
is then as follows:
\begin{enumerate}
\item Write a deterministic metainterpreter for chain programs
ignoring unification.
\item Incorporate to the metainterpreter
one of the transformations converting either moded or
unmoded programs into chain form.
\item In case the unmoded transformation was selected,
add renaming and unification to the metaclauses dealing with the
object {\em unit} clauses.
\end{enumerate}
We observed that even after adding unification, we need only be
concerned about substitutions at a single point of the chain-program
metainterpreter, whereas in a metainterpreter written directly
the unifiers are pervasive (cf.\ Fig.~\ref{clgr}).

\subsection{Performance study}
We have made a study illustrating how the performance of some programs is
degraded as a result of transforming such programs into chain form.
For this study, we used SICStus Prolog version 3.7.1, which we ran
under RedHat Linux version 6.0.
In this table and the next,
the columns labeled {\em A} show
the data for the source program and
the columns labeled {\em B} show
 the data for the corresponding transformed program.

First we exhibit the number of clauses and the 
program size (measured in bytes for compiled code).
\begin{center}
\begin{tabular}{lrrrr}\hline \hline
\multicolumn{1}{c}{\em program} & 
\multicolumn{2}{c}{\em num.\ clauses} & 
\multicolumn{2}{c}{{\em program size}} \\ 
& \multicolumn{1}{c}{\em A} & \multicolumn{1}{c}{\em B} 
& \multicolumn{1}{c}{\em A} & \multicolumn{1}{c}{\em B}
\\ \hline
{\it split} (moded {\it append}) & \hspace{15pt} 2 & 4 & 445 & 1,051
\\
{\it append} (for splitting) & 2 & 4 & 440 & 1,168
\\
{\it quicksort}, ord.\ lists & 7 & 22 & 1,405 & 5,833
\\
{\it quicksort}, diff.\ lists & 5 & 17 & 1,066 & 5,871
\\ \hline \hline
\end{tabular}
\end{center}

Now we give the relative execution times (according to SICStus'
\verb|profile_data/4|) 
and the memory requirements for the local and global stacks
(according to SICStus' \verb|statistics/0|)
for splitting a 100-element list into all its prefixes and suffixes
and sorting the reverse of a sorted 100-element list with quicksort programs.
\begin{center}
\begin{tabular}{lrrrr}\hline \hline
\multicolumn{1}{c}{\em program} & 
\multicolumn{2}{c}{\em execution time} & 
\multicolumn{2}{c}{{\em global+local stacks}} \\ 
& \multicolumn{1}{c}{\em A} & \multicolumn{1}{c}{\em B} 
& \multicolumn{1}{c}{\em A} & \multicolumn{1}{c}{\em B}
\\ \hline
{\it split} (moded {\it append}) & 26 & 571 & 32,760 & 32,760
\\
{\it append} (for splitting) & 26 & 636 & 32,760 & 32,760
\\
{\it quicksort}, ord.\ lists & 2,360 & 7,539 & 147,720 & 1,310,400 
\\
{\it quicksort}, diff.\ lists & 2,007 & 5,788 & 81,900 & 1,179,360 
\\ \hline \hline
\end{tabular}
\end{center}

\subsection{Related work}
Our work stemmed from the continuation-based and
the stream-based exhaustive-traversal methods.
There are, however, various other publications studying deterministic
traversals of search spaces within logic 
programming~\cite{hcf84,bst87,csh87,lcs87,shp87,stm89a,mdm93}.
Of these contributions,~\cite{mdm93} has perhaps the closest motivation
to ours:
The authors sketch a reconstruction of the continuation-based method
and give a metainterpreter of their exhaustive-search method (based on
recomputation).

Similarly, there are a number of transformations converting logic programs
into a syntactically restricted form~\cite{stm89,tby90,tr91}.
The one in~\cite{stm89} has in common with our work a
connection with the continuation-based exhaustive-traversal method.
These transformations differ from ours, however, in producing
programs in which every clause is binary (i.e.\ has only one atom in
the body).

\subsection{Future work}
We have argued that writing deterministic metainterpreters is
advantageous with our approach because 
this task amounts to that of describing an evaluation
strategy for relational expressions of the form: 
\begin{eqnarray*}
&& P = P_1 \cup P_2 \cup \ldots \cup P_m
\end{eqnarray*}
where each $P_i$ is defined as: 
\begin{eqnarray*}
&& P_i = Q_1 \comp Q_2 \comp \ldots \comp Q_n
\end{eqnarray*}
Evaluation strategies for  
expressions 
have a close connection with
the implementation of functional programming languages~\cite{pyt87} and
rewrite systems~\cite{djn90}.
Investigating how different evaluation strategies for these relational
expressions lead to different search
strategies for spaces determined by chain programs would be one way of
extending our contributions.

During the presentation of our results, we came across the need for
eliminating the layer of interpretation.
Our work would have a greater practical impact if it were combined with
an algorithmic elimination of such a layer, without producing
an excessively large residual program.

Chain programs have also proved to be useful in
devising~\cite{rsn96,rpr96}
inference systems derived from context-free parsers, 
because we need only consider unification in the treatment of unit
clauses and hence we need only modify the treatment of terminals.
Studying other applications of chain programs where it might be
helpful to relegate the role played by unifiers would be another
avenue of research.

\section*{Acknowledgments}
This work owes much to Carlos Velarde.
He contributed with motivating discussions, 
he carefully read previous versions of this paper, and
he spotted errors in the theorem proofs.
We also thank the referees, whose comments substantially improved the
presentation of these results.
We gratefully acknowledge the facilities provided by IIMAS, UNAM.

\bibliography{book,proc,article,tech,coll}
\end{document}